%% file: main.tex
\documentclass[sigconf,nonacm]{acmart}

\usepackage[labelfont=bf,font=small]{caption} 

\input{preamble_customizations}

\AtBeginDocument{%
  }

\copyrightyear{2026}
\acmYear{2026}
\setcopyright{cc}
\setcctype{by}
\acmConference[CHI '26]{Proceedings of the 2026 CHI Conference on Human Factors in Computing Systems}{April 13--17, 2026}{Barcelona, Spain}
\acmBooktitle{Proceedings of the 2026 CHI Conference on Human Factors in Computing Systems (CHI '26), April 13--17, 2026, Barcelona, Spain}
\acmPrice{}
\acmDOI{10.1145/3772318.3790622}
\acmISBN{979-8-4007-2278-3/2026/04}

\begin{document}

\title[Comparing How Students and Employees Perceive their University's LLM Chatbot vs. ChatGPT]{Campus AI vs. Commercial AI: Comparing How Students and Employees Perceive their University's LLM Chatbot vs. ChatGPT}


\author{Leon Hannig}
\orcid{0009-0007-9212-9037}
\affiliation{%
  \institution{University of Duisburg-Essen}
  \city{Duisburg}
  \country{Germany}
}
\email{leon.hannig@uni-due.de}

\author{Annika Bush}
\orcid{0000-0003-1475-4260}
\affiliation{%
  \institution{TU Dortmund University}
  \city{Dortmund}
  \country{Germany}
}
\email{annika.bush@tu-dortmund.de}

\author{Meltem Aksoy}
\orcid{0000-0003-3232-3923}
\affiliation{%
  \institution{TU Dortmund University}
  \city{Dortmund}
  \country{Germany}
}
\email{meltem.aksoy@tu-dortmund.de}

\author{Tim Trappen}
\orcid{0009-0004-4502-1777}
\affiliation{%
  \institution{Ruhr University Bochum}
  \city{Bochum}
  \country{Germany}
}
\email{Tim.Trappen@ruhr-uni-bochum.de}

\author{Steffen Becker}
\orcid{0000-0001-7526-5597}
\affiliation{%
  \institution{Ruhr University Bochum}
  \city{Bochum}
  \country{Germany}
}
\additionalaffiliation{%
  \institution{Max Planck Institute for Security and Privacy}
  \city{Bochum}
  \country{Germany}
}
\email{steffen.becker@rub.de}

\author{Greta Ontrup}
\orcid{0000-0003-4720-1494}
\affiliation{%
  \institution{University of Duisburg-Essen}
  \city{Duisburg}
  \country{Germany}
}
\email{greta.ontrup@uni-due.de}

\renewcommand{\shortauthors}{Hannig et al.}

\input{section/00_abstract}

\input{keywords}

\maketitle

\input{section/01_introduction}

\input{section/02_background}

\input{section/03_methods}
\input{section/04_results}

\input{section/05_discussion}

\input{acknowledgements}

\bibliographystyle{ACM-Reference-Format}
\bibliography{bibliography}


\input{section/07_Appendix}

\end{document}

%% file: preamble_customizations.tex
\usepackage{enumerate}
\usepackage{subcaption}
\usepackage{multirow}
\usepackage{rotating}
\usepackage{siunitx}

\usepackage{textgreek}
\usepackage{enumitem}
\usepackage{xspace}

\usepackage{svg}
\usepackage{dcolumn}
\usepackage[nolist]{acronym}
\usepackage{makecell}

\input{acronyms}

\setlist[enumerate,1]{label=(\arabic*)}
\setlist[enumerate,2]{label=(\theenumi.\arabic*)}
\setlist[enumerate,3]{label=(\theenumii.\arabic*)}

\setlist[enumerate,1]{label=(\arabic*), ref=\arabic*}
\setlist[enumerate,2]{label=(\theenumi.\arabic*), ref=\theenumi.\arabic*}
\setlist[enumerate,3]{label=(\theenumii.\arabic*), ref=\theenumii.\arabic*}

\newcommand{\ie}{i.\,e.}
\newcommand{\eg}{e.\,g.}

%% file: acronyms.tex
\begin{acronym}

    \acro{AI}{Artificial Intelligence}

    \acro{LLM}{Large Language Model}
    \acro{LLMaaS}{LLMs as-a-Service}

    \acro{TraM}{Trustworthiness Assessment Model}

\end{acronym}

%% file: section/00_abstract.tex
\begin{abstract}
As the use of \acf{LLM} chatbots by students and researchers becomes more prevalent, universities are pressed to develop AI strategies. 
One strategy that many universities pursue is to customize pre-trained LLM-as-a-service (LLMaaS) chatbots.
While most studies on LLMaaS chatbots prioritize technical adaptations, these systems are  often mainly characterized by user-salient front-end customizations, \eg, interface changes. Yet, no existing studies have examined how users perceive such systems compared to commercial LLM chatbots. In a field study, we investigate how students and employees ($N = 526$) at a German university perceive and use their institution's customized LLMaaS chatbot compared to ChatGPT. 
Participants using both systems ($n = 116$) reported greater trust, higher perceived privacy, and less perceived  hallucinations with their university's customized LLMaaS chatbot compared to ChatGPT. 
We discuss implications for research on users' trustworthiness assessment process, and offer guidance for the design and deployment of LLMaaS chatbots.

\end{abstract}

%% file: keywords.tex
\ccsdesc[500]{Human-centered computing~Empirical studies in HCI}
\ccsdesc[500]{Human-centered computing~User studies}
\ccsdesc[300]{Human-centered computing~Natural language interfaces}
\ccsdesc[300]{Security and privacy~Privacy protections}

\keywords{Large Language Models as-a-Service (LLMaaS), Customization, Trust, Hallucinations, University}

%% file: section/01_introduction.tex
\section{Introduction}
\label{sec::introduction}
\acf{LLM} chatbots are becoming increasingly prevalent in educational settings. Current research shows high adoption rates of commercial off-the-shelf LLM chatbots (\eg, ChatGPT) among students and university staff \cite{Zhang2024, Zhou2023, vonGarrel2023}. Consequently, universities face mounting pressure to develop strategies for AI integration \cite{kapania2024}. This involves multiple decisions, such as whether and how to provide employees with access to LLMs. 
To guide the informed and secure use of such systems, many organizations, including universities, corporations and government agencies, are exploring the deployment of LLM-as-a-Service (LLMaaS) as a feasible and cost-effective way to adopt AI in-house \cite{laMalfa2023, chkirbene2024, Odede2024}.
The adoption of such services is driven by several practical considerations, including the opportunity to offer access to all university members while maintaining institutional oversight and alignment with organizational policies \cite{paavola2024}. One of the main benefits of LLMaaS is the flexibility they offer in \textit{customizing} various aspects \cite{Diaferia2022}, such as fine-tuning models or displaying user interfaces in corporate design.

\textit{Customizations} of LLMaaS solutions play a critical role, as they can enhance model performance within specific organizational contexts and strengthen security and privacy measures \cite{sundberg2023, Lewicki2023}. 
However, even customization choices that have no impact on aspects such as the model or training data, but that are salient to users (\eg, in the form of cues perceivable in the user interface), could lead to significant changes in how the system is perceived and used. The focus of our work is on such front-end customizations, \eg, interface changes, and the role they may play in how users perceive and engage with the system.
Psychological research suggests that people draw inferences about a system’s trustworthiness based on indicators available to them, also referred to as \textit{cues} \cite{deVisser2014}, a mechanism formalized in the \textit{Trustworthiness Assessment Model} (TrAM) \cite{schlicker2025trustworthiness}. 
Real-world incidents show the importance of understanding user perception of LLM chatbots: cases like the data leak by Samsung employees via ChatGPT illustrate how false privacy expectations can lead to the unintended disclosure of sensitive information \cite{Mauran2023}. Such incidents underscore the importance of understanding user perceptions of differently embedded LLMs, \eg, LLMs offered as a Service vs. commercially available LLMs.

In this work, we examine if customized LLMaaS are perceived and used differently compared to commercially available alternatives. We draw on the TrAM \cite{schlicker2025trustworthiness} to theorize on the alignment (or misalignment) between users' perception of a customized LLMaaS and the system's actual characteristics, thereby answering the following research question:
\begin{list}{}{\leftmargin=0pt \rightmargin=0pt \itemsep=0.3\baselineskip \topsep=0.3\baselineskip}
\item
\fcolorbox{green!75!black}{green!5!white}{\parbox{0.95\linewidth}{\textbf{RQ:} How do users' trust, perception, and usage differ comparing a university's customized LLMaaS chatbot and a commercially available LLM chatbot?}}
\end{list}
To examine this research question, we conducted a survey-based field study at a major German university. A diverse sample of students and university employees (N = 526) answered questions about their experiences with the university’s customized LLMaaS chatbot and ChatGPT (the most widely used LLM chatbot among German students \cite{vonGarrelMayer2025}). We measured trust, privacy concerns, cautious behavior toward LLM hallucinations, perceived LLM hallucinations, and sustainability-aware AI usage. Our main analyses focus on a within-subject subsample of participants (n = 116) who regularly used both systems, allowing direct comparison of user perceptions and behaviors across the two platforms.

Our primary contributions are:
\begin{enumerate}
    \item We show that participants indicate greater trust, higher perceived privacy and less perceived LLM hallucinations with their university's customized LLMaaS chatbot compared to ChatGPT.
    \item Drawing on the TrAM framework \cite{schlicker2025trustworthiness}, we theorize that customizations of LLMaaS serve as cues that could be the reason for differences in users' perceptions and behavior toward LLMaaS chatbots compared to commercially available LLM chatbots.
    \item We extend research on calibrated trust by broadening the concept of calibration to include privacy and hallucinations, demonstrating their relevance in an actual academic LLMaaS use-case.
    \item We provide organizations and developers with practical recommendations regarding the deployment and customizations of LLMaaS chatbots to support their informed, safe and sustainable use.
\end{enumerate}

%% file: section/02_background.tex
\section{Theoretical Background and Hypotheses}
\label{sec::background}

\subsection{LLM (as-a-Service) Chatbots at Universities}
\label{subsec::background::LLMs_at_Universities}
LLM chatbots, like ChatGPT, are widely used in academic settings, especially among students \cite{vonGarrel2023, amani2023} and researchers \cite{kapania2024, syed2024awareness, abdelhafiz2024knowledge}. 
Thus, universities are caught between enabling students and staff to access and use AI in an informed way to leverage the benefits while also protecting data, privacy, rights, and ethical principles. 
Introducing ``their own'' LLMs poses a considerable challenge for universities, due to a lack of internal technical abilities and knowledge about the deployment of AI and the high costs of developing and maintaining the required IT infrastructure \cite{lins2021}. 

LLMs-as-a-Service (LLMaaS) provide a practical solution by offering easy access to pre-trained models via cloud providers \cite{wang2024, syed2025}. 
This approach is a subset of AI-as-a-Service (AIaaS), broadly defined as ``cloud-based systems providing on-demand services to organizations and individuals for deploying, developing, training, and managing AI models'' \citep[p. 424]{lins2021}. 
These systems enable organizations to adopt and afford AI technologies without requiring extensive technological expertise \cite{sundberg2023, syed2025}. 
Moreover, LLMaaS solutions emphasize enhanced data protection and security, offering a viable option for institutions like universities to leverage AI's advantages while adhering to stringent compliance and security standards \cite{paavola2024}. 

\subsection{Customization of LLMaaS}
One proposed advantage of LLMaaS is that organizations can customize various aspects of LLMs to better align with their specific requirements and objectives.
Such customizations span different architectural layers -- (i)~the front-end and (ii)~the back-end, which comprise most chatbot applications, (iii)~the LLMaaS API, and (iv)~the underlying models -- to improve the fit between a system and organizational needs \cite{Diaferia2022}.
Within the broader AI supply chain \cite{cobbe2023}, these modifications may be implemented transparently or remain obscure to end users \cite{neumann2025a}.
Although LLMaaS architectures permit customization across all layers, their implications differ markedly.
Back-end, API, or model-level customizations primarily affect technical performance, interoperability, or governance, whereas front-end customizations -- the layer directly encountered by users -- constitute the most salient and psychologically consequential touchpoints.

Existing research has predominantly examined technical customizations, \eg, model fine-tuning \cite{arun2024, park2025}, integration and interoperability \cite{Weber2024, Torres2025}, or security and governance considerations \cite{Cai2025, Weber2024}.
In contrast, user-facing customizations remain comparatively underexplored, despite their potential impact on user perceptions, trust, and behavior. These front-end modifications may include transparency features (\eg, explainability panels \cite{gong2024}) or visual branding (\eg, logos, color schemes) to align with institutional identity \cite{Casati2019}.

Because branding is frequently intertwined with customization in both practice and scholarly discourse, some conceptual disentangling is necessary.
In marketing research, customization is often positioned within broader branding strategies \cite{Yan.2020}, yet organizational AI deployment demands a clearer distinction.
Branding establishes a recognizable identity through visual or symbolic elements \cite{Bertram.2016}, while customization denotes the wider set of adaptations tailored to institutional needs.
Branding-related modifications thus represent only one subset of user-facing customizations. 
We therefore use the broader term ``customization'' to capture the full range of user-facing adaptations in LLMaaS deployments, while recognizing branding as one particularly visible -- but not exclusive -- customization strategy.

In this work, we compare a customized LLMaaS chatbot -- primarily modified at the front-end, with only minimal back-end adjustments -- to a commercially available alternative (ChatGPT).

\subsection{System Characteristics and User Perceptions: The Role of Cues}
The aim of this work is to investigate whether users' perceptions and use of customized LLMaaS chatbots differs from commercially available LLM chatbots. 
The TrAM model by \citet{schlicker2025trustworthiness} allows us to theorize on why there might be differences in perceptions and therefore serves as a basis to derive our hypotheses.
The TrAM posits that users assess the actual trustworthiness of a system based on available cues, leading to the perceived trustworthiness \cite{schlicker2025trustworthiness}. Thus, users' trust and corresponding use of AI depend on, \ie, ``any information element that can be used to make a trust assessment about an agent'' \citep[p. 253]{deVisser2014}. The accuracy of this assessment relies on the relevance and availability of cues on the system's side, as well as the human's ability to detect and utilize them. On the one hand, actual system characteristics themselves can function as cues, for example information about the training data or the system’s outputs. On the other hand, users may also rely on cues that are less directly tied to a system’s actual qualities, such as interface design, which can still strongly influence their perceived trustworthiness \cite{gong2024}. Such cues may therefore induce unwarranted trust (or distrust) and lead users to perceive systems as more (or less) trustworthy than their actual reliability would justify.

The balance or imbalance between perceived vs. warranted trust is also discussed under the umbrella term 'trust calibration'. This concept describes that users' perceived trustworthiness should accurately match the actual capabilities and reliability of the system \cite{lee2004trust}. \textit{Miscalibration} can manifest as either \textit{undertrust}, leading to system disuse, or \textit{overtrust}, resulting in system misuse (\eg, failures of monitoring or decision biases) \cite{ParasuramanRiley1997, WischnewskiKraemerMueller}. The fit between the actual trustworthiness of a system and its perceived trustworthiness therefore represents one facet of trust calibration (referred to as ‘high accuracy of the trustworthiness assessment’ in the TrAM model; \cite{schlicker2025trustworthiness}). 

We argue that the concept and relevance of the trustworthiness assessment and the resulting calibration applies not only to trust, but to user perceptions in general. 
We propose to conceptualize the trustworthiness assessment process with regards to user perceptions in the context of LLM use more broadly and to consider not only users' trust as an important outcome, but also the perceived privacy, hallucinations as well as sustainable use of the system. Building on the TrAM logic of linking actual system characteristics to perceived system characteristics via cues that are salient to the user, allows us to conceptualize whether there is a fit between actual system features (\eg, hallucination tendency, privacy safeguards) and end users' perceptions of those corresponding system features (\eg, perceived hallucinations or perceived privacy). We interpret this fit as the degree of calibration of user perceptions with regards to actual system characteristics.

\begin{figure*}[ht]
\centering
\includegraphics[width=0.9\textwidth]{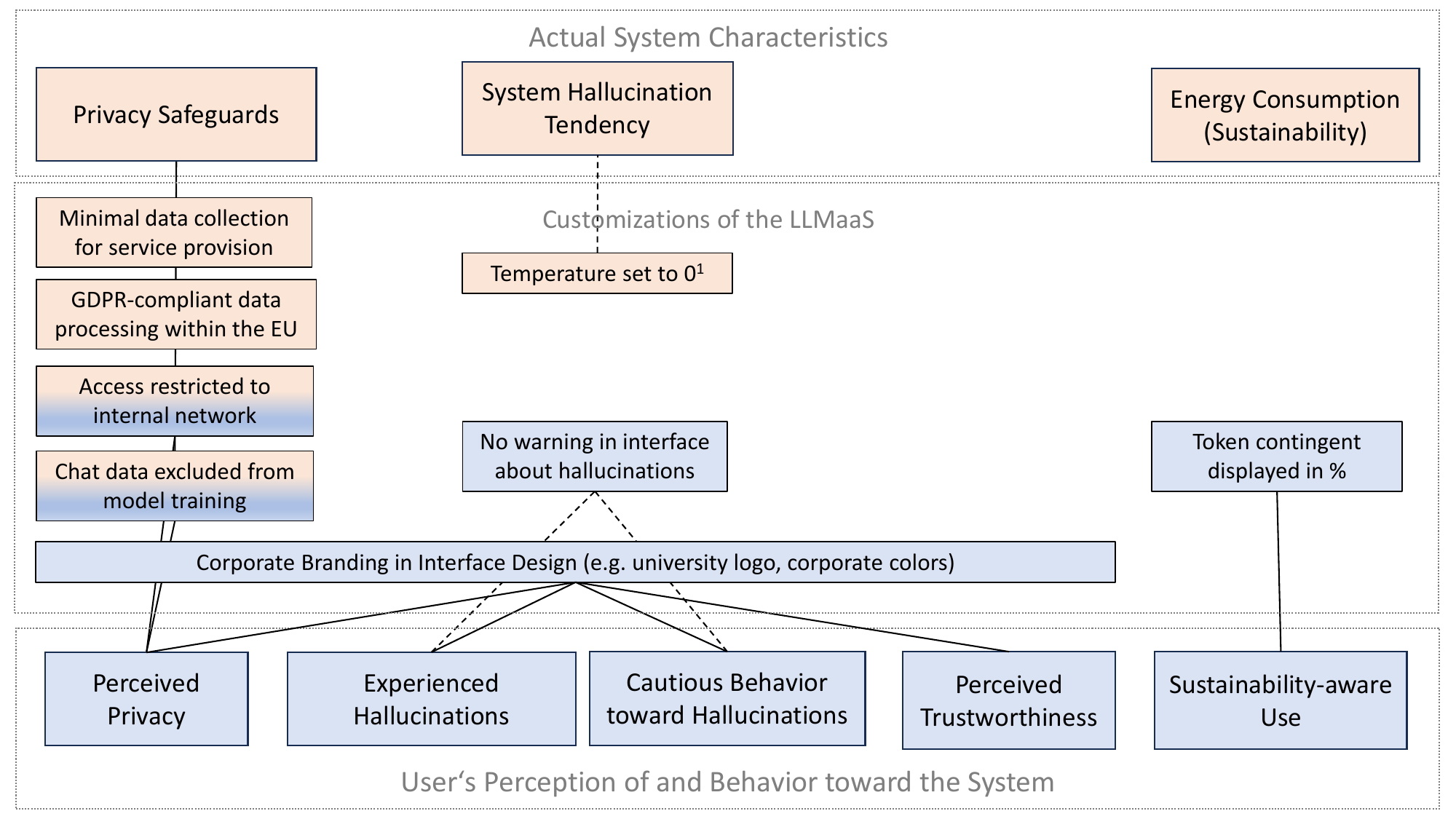}
\caption{\textbf{Theoretical foundation: System characteristics, customization cues, and user perceptions for the examined university's customized LLMaaS chatbot.} Solid lines represent an expected positive association between the construct, dotted lines represent an expected negative association. Customizations colored in red relate to \textit{actual} system characteristics, while we expect those in blue to solely relate to \textit{perceived} characteristics. Red-and-blue customizations are expected to have associations with both. Footnotes: ¹Lower temperature may reduce the hallucination tendency by favoring higher-probability tokens and making the output more deterministic \cite{Huang_2025}. 
}
\Description{Diagram showing theoretical basis to expected relationships between actual system characteristics, customization cues, and user perceptions for the university’s customized LLMaaS chatbot. 
Solid lines indicate positive associations between constructs, and dotted lines indicate negative associations. 
The diagram is organized into three horizontal sections.

Top section – Actual System Characteristics: This part contains three main categories: Privacy Safeguards (red), System Hallucination Tendency (red), and Energy Consumption – Sustainability (red).

Middle section – Customizations and Cues:
Under Privacy Safeguards (red), four rectangular boxes are listed vertically: “Minimal data collection for service provision” (red), “GDPR-compliant data processing within the EU” (red), “Access restricted to internal network” (red-and-blue), and “Chat data excluded from model training” (red-and-blue).

Under System Hallucination Tendency (red), the smaller box “Temperature set to 0” (red) is connected with a dotted line to the main category. 
Below it, “No warning in interface about hallucinations” (blue) is linked with dotted lines to perceived Hallucinations and Cautious Behavior toward Hallucinations.

Under Energy Consumption – Sustainability (red), the box “Token contingent" displayed in blue is connected downward to Sustainability-aware Use.

One wide box, “Corporate Branding in Interface Design (\eg, university logo, corporate colors)” (blue), spans the middle and bottom sections and is connected to all user perception and behavior constructs except Sustainability-aware Use.

The bottom section, named "User’s Perception of and Behavior toward the System", contains five boxes in a row, all blue: Perceived Privacy, perceived Hallucinations, Cautious Behavior toward Hallucinations, Perceived Trustworthiness, and Sustainability-aware Use.

The following connections are symbolized through lines:
Red-only customizations link only to Actual System Characteristics.
Blue-only customizations link only to User’s Perceptions and Behaviors.
Red-and-blue customizations (“Access restricted to internal network” and “Chat data excluded from model training”) connect to both Actual System Characteristics and User’s Perceptions and Behaviors.}

\label{fig:TrAM_Figure}
\end{figure*}

It is essential to recognize that end users could perceive customizations of LLM chatbots as reflections of the system's capabilities, goals, and inner workings, even if customizations do not actually affect the functioning of the system. \autoref{fig:TrAM_Figure} illustrates our theoretical reasoning with regards to the expected interplay between customizations (partially functioning as cues) of the LLMaaS chatbot in relation to actual system characteristics and user perceptions or self-reported behaviors. Importantly, this illustrates the basis for the derivation of theoretically-grounded differential hypotheses (see following chapters). These are not mechanisms of action that we explicitly test in an experimental setup. In the following, we elaborate on the theoretical foundations of these expected differences in perception in relation to specific constructs.

\subsection{Trust in AI Systems}
\subsubsection{Background}
Trust, as a key determinant of technology adoption \cite{van2019trust, Choudhury2023}, has been recognized as an important factor for LLM chatbot adoption \cite{bhaskar2024shall, Tiwari2024}.
Trust is commonly defined as the trustor's positive expectations towards the trustee, combined with the trustor's vulnerability and uncertainty within the specific context \cite{balayn2024, lee2004trust}. 
Research has shown that trust levels in human-human interaction depend on three perceived qualities of the trustee: ability, integrity, and benevolence \cite{mayer1995}. 
In the context of human-automation interaction, these three factors have been reframed as performance, process (\eg, availability, confidentiality, understandability of a system) and purpose (\eg, the developer's or deployer's intentions) \cite{lee2004trust}. 
Additional factors have been identified as particularly important for trust in LLMs. 
These include transparency \cite{huang2023, schwartz2023}, access to a human operator \cite{nordheim2019}, perceived privacy and robust data security \cite{huang2023, pawlowska2024}. 

\subsubsection{Customizations and Hypotheses}
We theorize that users report different levels of trust in a customized LLMaaS chatbot compared to a commercially available LLM chatbot. Drawing on the TrAM, we expect customizations of the LLMaaS user interface to serve as a salient trust cue for users, increasing the perceived trustworthiness of the system in two dimensions: (a) purpose and (b) process. 
(a) The perceived purpose of the LLMaaS system may appear more trustworthy when university branding is incorporated into the front-end of the chatbot (see \autoref{fig:User_Interface}, Label~1)\footnote{\autoref{fig:User_Interface} shows a replica of the LLMaaS chatbot's user interface, with customizations marked in red.}, as the university is likely to be seen as a benevolent provider with minimal financial or marketing-driven motives \cite{nordheim2019}. In this context, university branding refers to the integration of the university’s logo and color scheme into the user interface, creating a visually prominent link between the system and the institution. 
The incorporation of university branding in the front-end may also create a sense of familiarity, a factor found to foster interpersonal trust and that might also support trust in AI when users associate it with a familiar institution \cite{Alarcon2016}. 
(b) The process dimension of the system may be perceived as more trustworthy because it is easily accessible and embedded in the existing organizational framework. 
Here, features such as the option to contact a human operator for assistance could contribute to users' trust through perceived openness and accessibility of the system \cite{lee2004trust}. 
Additionally, the customized LLMaaS chatbot may be perceived as more transparent due to context-specific information about the system available on the university's website. 
These factors together lead to the hypothesis: 
\begin{list}{}{\leftmargin=0pt \rightmargin=0pt \itemsep=0.3\baselineskip \topsep=0.3\baselineskip}
\item
\fcolorbox{blue!75!black}{blue!5!white}{\parbox{0.95\linewidth}{\textbf{H1:} Users report higher levels of trust in the customized LLMaaS chatbot compared to a commercial LLM chatbot.}}
\end{list}

We further hypothesize that higher trust in the organization's customized LLMaaS chatbot will likely depend on university members' trust in the university itself. In reference to the TrAM, we consider organizational trust to be an external, moderating factor that has an impact on the association between perceived system characteristics and reported trust in a system.
Organizational trust describes the extent to which individuals trust an organization~\cite{Pirson2011}. 
Prior research suggests that organizational trust can significantly shape users' trust in the organization's system \cite{balayn2024, Lapinska2021}. 
In line with this, \citet{nordheim2019} have demonstrated that customers' trust in a company's service chatbot is influenced by their trust in the company itself. A similar mechanism might apply in the context of academic institutions: users’ trust in a university-provided LLMaaS chatbot may be influenced by their trust in the university. 
We therefore hypothesize:
\begin{list}{}{\leftmargin=0pt \rightmargin=0pt \itemsep=0.3\baselineskip \topsep=0.3\baselineskip}
\item
\fcolorbox{blue!75!black}{blue!5!white}{\parbox{0.95\linewidth}{\textbf{H2:} The level of trust in the university moderates the effect of system type (customized vs. commercial) on user trust, such that the effect is stronger when organizational trust is higher and weaker when it is lower.}}
\end{list}

\subsection{LLM Hallucinations} 

\subsubsection{Background}
Hallucinations in LLMs refer to the generation of content that appears plausible but is factually incorrect or inconsistent with user input \cite{Huang_2025}. 
Hallucinations in human-AI interaction pose a dual challenge: they impair effective collaboration by reducing output quality and undermining users' trust \cite{gong2024, Amaro2024}. 
In the academic context, hallucinations can pose particular challenges, as they can reduce perceived accuracy and usefulness, increase confusion, and prolong task completion~\cite{Steinbach2025}.
Such issues are compounded by the fact that LLMs have been shown to often fabricate inaccurate scientific references~\cite{Aljamaan2024, Chelli2024}.

\subsubsection{Customizations and Hypotheses}
\begin{figure*}[h!]
\centering
\includegraphics[width=0.7\textwidth]{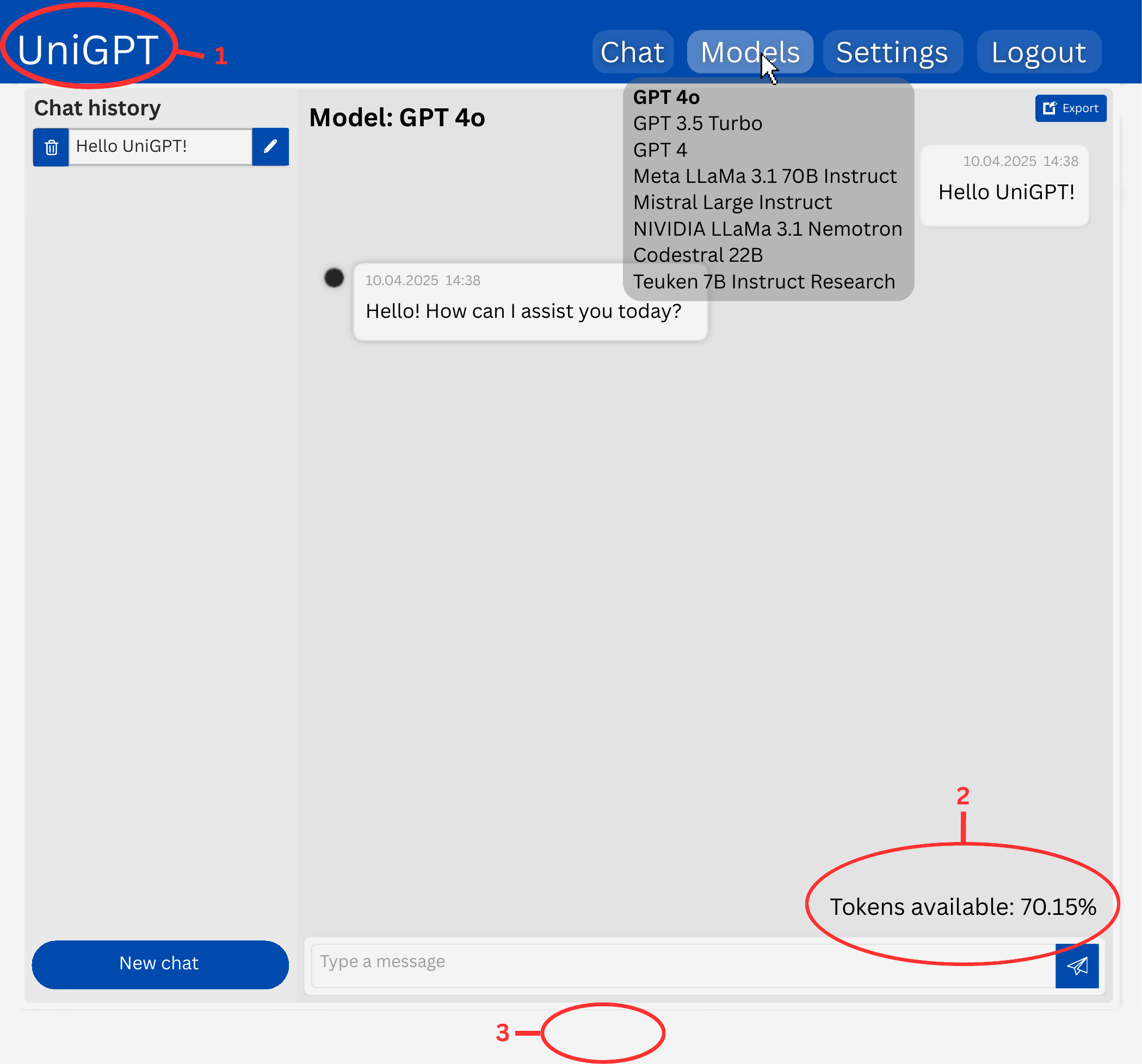}
\caption{\textbf{Anonymized replica of the user interface of the university’s customized LLMaaS chatbot.} Key elements: (1) university name is displayed; interface follows university's visual design; (2) remaining monthly token contingent shown in percent; (3) no warning regarding potential hallucinations / incorrect outputs.}
\Description{Screenshot of the anonymized replica of the university’s customized LLMaaS chatbot interface. 
The top navigation bar is blue with four buttons on the right: Chat, Models, Settings, and Logout. 
On the left, the label “UniGPT” is displayed, styled to match the university’s visual identity (1). 
Beneath the navigation bar is the chat area. On the left panel, a chat history lists a single entry “Hello UniGPT!”. 
The main panel shows a conversation with a greeting from the chatbot (“Hello! How can I assist you today?”). 
A drop-down menu labeled “Models” is open, listing model options such as GPT-4o, GPT-3.5 Turbo, GPT-4, Meta LLaMa, Mistral Large Instruct, NVIDIA LLaMa, Codestarl, and Teuken 7B. 
On the right side of the main panel, a status element (2) displays the remaining monthly token contingent as a percentage (“Tokens available: 70.15\%”). 
Below the chat window is a text entry field with a send button on the right. 
Area (3) is an empty space, highlighting the absence of any visual warning about potential hallucinations or incorrect outputs in the interface.}
\label{fig:User_Interface}
\end{figure*}

We expect there to be differences with regard to users' perception of and cautious behavior towards hallucinations and ground this expectation in two customization aspects: branding through corporate design and warnings about false information. Thus, with reference to the TrAM, we conceptualize branding with corporate design and warnings about false information as system cues and consider users' perception of and cautious behavior towards hallucinations as important 'outcome' perception dimensions.
Research has shown that the presence of logos from reputable companies can create positive associations, enhancing the perceived credibility of mobile systems \cite{lowry2005}. 
We argue that branding a chatbot with a university logo (see \autoref{fig:User_Interface}, label 1) leads users to interpret the system as university-approved. 
This could potentially make users less critical and cautious in their interactions with the university-branded system, particularly with regard to the risk of hallucinations.
Another notable customization choice of LLMaaS chatbots is the display of warnings about hallucinations or false information. 
Research suggests that such warnings can help users detect hallucinations more frequently \cite{nahar2024}. 
While ChatGPT's interface includes a disclaimer below the prompt box stating ``ChatGPT can make mistakes. Please check important information'', LLMaaS solutions allow such warnings to be adapted, removed, or altered. In our study, the customized LLMaaS chatbot included a hallucination disclaimer only on the login page, with no warning displayed during the actual chat interaction (see \autoref{fig:User_Interface}, label 3).
We argue that the absence of a persistent warning may result in users being less mentally primed to consider hallucinations or to critically evaluate the accuracy of the system's output.
The university-branded interface and the lack of a persistent warning may lead users to a) behave less cautiously with respect to potential hallucinations and b) detect fewer hallucinations overall.
Based on this, we propose the following hypotheses:

\begin{list}{}{\leftmargin=0pt \rightmargin=0pt \itemsep=0.3\baselineskip \topsep=0.3\baselineskip}
\item
\fcolorbox{blue!75!black}{blue!5!white}{\parbox{0.95\linewidth}{\textbf{H3:} Users report less cautious behavior towards hallucinations when using the customized LLMaaS chatbot compared to a commercial LLM chatbot.}}
\item
\fcolorbox{blue!75!black}{blue!5!white}{\parbox{0.95\linewidth}{\textbf{H4:} Users report fewer perceived hallucinations when using the customized LLMaaS chatbot compared to a commercial LLM chatbot.}}
\end{list}
To assess whether actual differences in hallucination tendencies between the systems might contribute to differences in users’ perceived hallucinations, we evaluate both systems using benchmark datasets (see \autoref{Exploratory Analyses}).

\subsection{Data Security and Privacy Concerns}
\subsubsection{Background}
End users often express concerns about data security and privacy when interacting with LLMs~\cite{Ma2025, Mutahar2025}. 
Reflecting these concerns, recent research has called for privacy-friendly user interfaces of LLM systems~\cite{Li_Tianshi2024, Song2025, Windl2025}.
Such concerns are grounded in tangible risks, including (unauthorized) data retention~\cite{cobbe2021}, where providers may retain and use customer data to enhance their models, inference attacks that aim to extract sensitive information from trained models~\cite{kandpal-etal-2024-user}, and the unintended exposure of confidential data during model training or operation~\cite{chen2024janus}.
To mitigate these risks, technical customization options in LLMaaS offer promising avenues, playing a pivotal role in enhancing users’ perceptions of privacy and security while fostering trust. These options include user-configurable privacy settings that empower users to control how their data is stored and used~\cite{Ashley2002}, transparency mechanisms like privacy explanations~\cite{BRUNOTTE2023111545}, which provide clarity about data handling practices, and the implementation of privacy-preserving techniques such as differential privacy~\cite{abadi2016deep} during training and fine-tuning processes.

\subsubsection{Customizations and Hypothesis}
In the LLMaaS chatbot under study, we identified two system characteristics with potential relevance for users’ privacy perceptions: restricted access to internal networks and the exclusion of chat data from model training (see \autoref{fig:TrAM_Figure}). In reference to the TrAM, these represent actual system characteristics, which users \textit{might} be able to perceive/ inform themselves about.
Beyond these direct technical benefits, we posit that other customization measures -- even those without tangible impacts on privacy or security -- may foster higher perceived privacy levels. In regard to the TrAM, we conceptualize those as perceived system characteristics that relate to users' privacy concerns but \textit{not} to the actual systems' privacy settings.
In this context, we again draw on the role of corporate branding in the interface design as a salient customization feature:
University branding (see \autoref{fig:User_Interface}, label 1) may enhance users perceptions of data security and privacy, as they may believe that their data will be handled with care and not shared with third parties. 
Prior research supports this assumption, showing that a strong brand image or high brand credibility can reduce perceived privacy risks and increase users’ willingness to disclose personal information online \cite{wang2019, myerscough2008, jain2022}.
Thus, we propose:
\begin{list}{}{\leftmargin=0pt \rightmargin=0pt \itemsep=0.3\baselineskip \topsep=0.3\baselineskip}
\item
\fcolorbox{blue!75!black}{blue!5!white}{\parbox{0.95\linewidth}{\textbf{H5:} Users perceive greater privacy when using the customized LLMaaS chatbot compared to a commercial LLM chatbot.}}
\end{list}

\subsection{Sustainable AI Use}
\label{subsec::background::sustainable}
\subsubsection{Background}
In this final part, we broaden the perspective to considerations that pertain to the global strategy of an organization, specifically the corporate social responsibility (CSR). 
CSR refers to situations in which organizations pursue strategies that go beyond business needs and legal regulations but aim at furthering social good~\cite{McWilliams2001}, \eg, improving an organization's environmental performance~\cite{McWilliams2006}.
We argue that customization of LLMaaS chatbots offers organizations opportunities to align AI deployment with their CSR strategy and corresponding sustainability goals, thereby not only shaping the trust in and perception of the systems but also their environmental impact.

The environmental impact of AI systems and LLMs has become an increasingly critical concern in computing research. 
Recent work by \citet{Rafael.2024} emphasizes how the invisibility of digital energy consumption often leads to unconscious and potentially wasteful usage behavior. 
Our research builds on these insights and considers the role of customization choices aimed at promoting sustainability-aware AI usage.

\subsubsection{Customizations and Hypothesis}
Making resource consumption visible has shown promising results across different domains. \citet{Penkert.2023} demonstrated that transparency in sustainability information significantly influences user choices, while \citet{Sanduleac.2017} found that real-time energy data promotes more responsible consumption patterns. 
LLMaaS chatbots offer customization options to make AI resource limitations explicit through token visibility (see \autoref{fig:User_Interface}, label 2). 
We argue that customizing token visibility can foster what we term ``sustainability-aware AI use'' -- a more deliberate approach to AI interaction that considers broader environmental impact. In reference to the TrAM, the displayed tokens represent perceived system characteristics that are salient to the user. We theorize that those might lead to differences in self-reported sustainability aware AI use.
Based on these previous findings, we hypothesize: 
\begin{list}{}{\leftmargin=0pt \rightmargin=0pt \itemsep=0.3\baselineskip \topsep=0.3\baselineskip}
\item
\fcolorbox{blue!75!black}{blue!5!white}{\parbox{0.95\linewidth}{\textbf{H6:} Users report higher levels of sustainability-aware AI usage when using the customized LLMaaS chatbots compared to a commercial LLM chatbots.}}
\end{list}

%% file: section/03_methods.tex
\section{Method}
\subsection{Study Design}
We conducted a quantitative, cross-sectional field study. Data was collected through an online survey (3rd March - 12th May 2025) at a major public university that had implemented a customized LLMaaS chatbot at the end of 2024. Participants who reported regular usage of both chatbots answered the same questions for the university's customized LLMaaS chatbot and for ChatGPT.

\subsection {Description of the University's Customized LLMaaS Chatbot}
The LLMaaS chatbot introduced by the university serves multiple LLMs hosted by different providers. The default model for each chat session, GPT-4o, as well as other OpenAI models, are provided by Microsoft Azure using a datacenter in Sweden. Furthermore, users can choose from a small selection of open-weight models provided by a German AI service center (see \autoref{fig:User_Interface}). Thus, only application layers (i) front-end and (ii) back-end are under the university's control. The system had 4636 users at the beginning of the data collection period and grew by $20.7\%$ to 5593 users by the end of the period.
Customizations regarding the back-end were kept to a minimum, but the university made some adjustments at the front-end, specifically to the user interface and incorporated detailed information (see \autoref{tab:customized_LLM}). In contrast to OpenAI’s ChatGPT, the university chatbot supports only text-based interactions and image generation.  \autoref{fig:User_Interface} shows an anonymized replica of the user interface.

\begin{table*}[ht]
\centering
\captionsetup{justification=raggedright,singlelinecheck=false}
\caption{\textbf{Customization features of the studied LLMaaS chatbot.}}
\label{tab:customized_LLM}
\small
\setlength{\tabcolsep}{4pt}
\renewcommand{\arraystretch}{1.15}
\begin{tabular}{p{0.30\textwidth} p{0.70\textwidth}}
\toprule
\textbf{Category} & \textbf{Customization Details} \\
\midrule
Model Fine-tuning and Configuration &
\parbox[t]{\hsize}{No fine-tuning or other model customizations. Temperature\textsuperscript{1} set to 0 for deterministic outputs. System prompt fixed to: "\textit{You're a helpful assistant.}" for all models, not modifiable by users} \\
\midrule
Integration and Interoperability &
Access to multiple models from different LLMaaS providers; image generation via DALL-E for OpenAI models \\
\midrule
Security and Governance &
Minimal data collection for service provision; chat data excluded from model training; GDPR-compliant data processing within the EU; access restricted to internal networks (\eg, via VPN) \\
\midrule
User Interface and Experience &
\parbox[t]{\hsize}{Custom-built interface following corporate design (\eg, university logo); monthly token contingent visualized as a percentage in the user interface; no warning about hallucinations at the bottom section; dropdown menu for choosing active model} \\
\midrule
Organizational Integration &
Supporting materials provided on the chatbot landing page and intranet; training sessions on effectively using LLMs in higher education \\
\bottomrule
\end{tabular}

\vspace{2mm}
\begin{minipage}{\textwidth}
\raggedright\small
$^{1}$ The temperature parameter in LLMs, ranging from 0 to 1, controls response randomness, with values near 0 favoring deterministic outputs and values near 1 producing diverse, creative outputs.
\end{minipage}
\end{table*}

\subsection{Ethical Considerations}
By involving the a) data protection officer, b) work councils, c) center for teaching and learning and d) information technology department, the rights and interests of all stakeholders were taken into account in the design of the study. The approval of all parties and the local ethics committee was sought prior to implementation. 
Concerning data collection, storage, and sharing, we followed the guidelines of the university's IT committee and used committee-approved survey software with university-internal data storage, ensuring that only the project team had access to the raw data.
The participants were informed about the content and procedure of the study via standard written instructions \cite{Schoenbrodt2017}. Participation was voluntary, anonymous, and uncompensated.

\subsection{Survey Structure and Measures}
\begin{figure*}[htbp]
\centering
\includegraphics[width=0.95\textwidth]{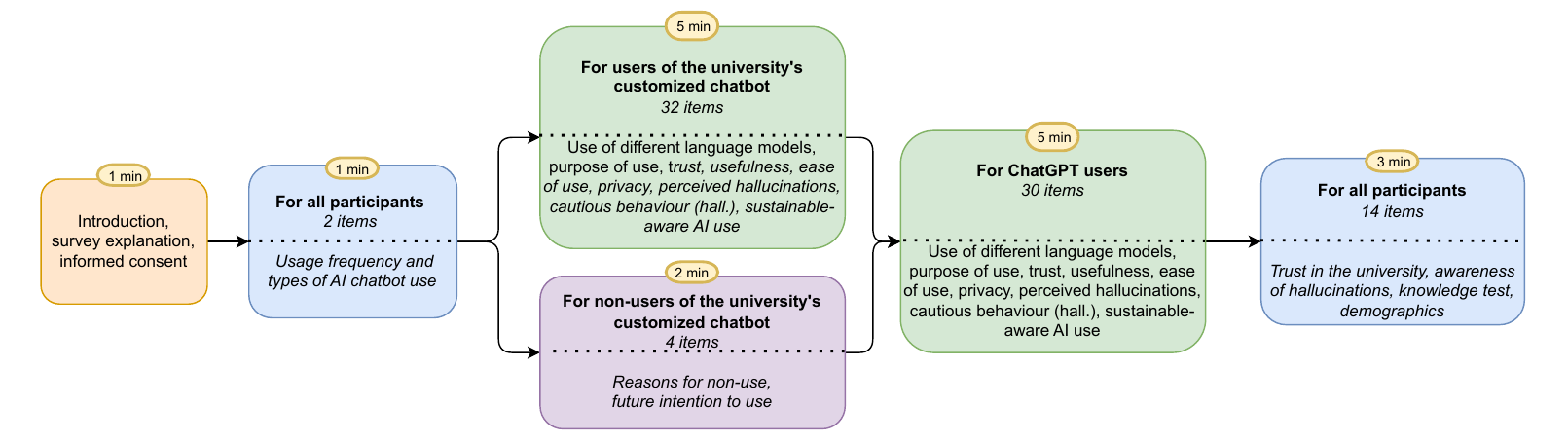}
\caption{\textbf{Survey flowchart illustrating participant routing and item blocks based on chatbot usage.} Labels at the top of each block indicate the estimated completion time for that section.}
\Description{Flowchart with six labeled blocks, connected by directional arrows, representing the survey structure for different participant groups. 
The sequence begins with an introductory module (“Introduction, survey explanation, informed consent”, 1 min). 
Next, all participants complete a block with two items on chatbot usage frequency and types of AI chatbot use (1 min). 
The flow then branches based on chatbot usage: participants complete either 
(a)~a 32-item module for users of the university’s customized chatbot (5 min), or 
(b)~a four-item module for non-users of this chatbot (2 min). 
Participants who report using ChatGPT additionally complete a 30-item module (5 min). 
All participants finish with a final 14-item module that covers trust in the university, general hallucination awareness, the knowledge test, and demographics (3 min).}
\label{fig:Survey_Flow}
\end{figure*}

We designed a comprehensive survey to investigate the perceptions and usage of both the university's customized LLMaaS chatbot and OpenAI's ChatGPT. ChatGPT was chosen as the commercial LLM because of its widespread use in Germany \cite{vonGarrelMayer2025}. Our data support this choice: in our sample, 333 participants reported using ChatGPT at least once a month, compared to 176 who reported comparable usage of other AI chatbots. During the data collection phase, ChatGPT was powered by OpenAI’s GPT-4o.\footnote{See \url{https://help.openai.com/en/articles/6825453-chatgpt-release-notes} for a detailed record of ChatGPT's development during our data collection period.}
The survey design is summarized in \autoref{fig:Survey_Flow}. 
Participants were guided through the survey based on their use/non-use of the chatbots. 
The main part consists of questions regarding the perception and use of the university chatbot and ChatGPT (green blocks in \autoref{fig:Survey_Flow}). An overview of all survey questions can be found in the supplementary material (see \autoref{open_science}).

All hypothesis-relevant items followed the same response format to minimize cognitive load and participant fatigue. We also conducted pilot runs to ensure that the average completion time remained under 15 minutes. Only hypothesis-relevant items were mandatory and were placed toward the beginning of the questionnaire to reduce the likelihood of fatigue affecting key measures.


The survey began by capturing usage patterns for both systems: frequency of use (ranging from ``never'' to ``several times a day'') and context of use (6-point scale in response to the item ``I tend to use the following AI chatbots for …'', ranging from 1 = ``private purposes'' to 6 = ``study-/work-related purposes'').
Next, several aspects of system perception and use were explored: specific use purposes (\eg, information search), perceived usefulness (productivity enhancement, task facilitation), and ease of use (interface clarity, learning curve); the latter two were assessed following the Technology Acceptance Model (TAM)~\cite{davis1989}.
For users familiar with either system, we also assessed their awareness of model switching options and use of different models. 
For users who have never or rarely used the university's LLMaaS, we explored their reasons for non-use.

At the end of the survey, a knowledge test was administered to assess participants' understanding of the customized LLMaaS chatbot. The test consisted of five items (\eg, ``The university chatbot offers models trained by the university'' (false)) with response options ``true'', ``false'', and ``I don't know''. Correct answers were summed to create a knowledge score ranging from 0 to 5. Participants were debriefed on the following page about the correct answers. At the end of the study, participants could voluntarily report demographic information.

The main part of the survey was structured around the key constructs. 
Participants who used both systems responded to the same set of items twice: once for the university's customized LLMaaS chatbot and once for ChatGPT. Only participants who reported using a system at least once a month were presented with the full set of items.
Participants were reminded to answer the questionnaire items with regards to their work-related or academic use of the system. 
We used five-point Likert scales ranging from ``strongly disagree'' to ``strongly agree'' for all scales. The questionnaire was available in German and English, with items from previously published scales translated for the German version. Internal consistencies (Cronbach’s $\alpha$) for the scales are reported in \autoref{tab:llmaas_correlations} and \autoref{tab:chatgpt_correlations}.

\paragraph{Trust (in the System)} Trust in the systems was assessed by the facets of global trust, benevolence/purpose, integrity/process, and ability/performance using four items from \citet{wischnewski2025} (\eg, ``I trust [the university chatbot/ChatGPT].''), forming an overall trust measure. We selected this scale for its recent validation, context specificity, and strong fit to human–AI interaction contexts, as well as its theoretical grounding in calibrated trust research \citep{lee2004trust, ParasuramanRiley1997}.

\paragraph{Trust in the university}
Trust in the university was measured with three items following the \textit{trust in management} scale by \citet{Boe2018} (\eg, ``My university will safeguard my interests.''). We selected this scale due to its strong psychometric validation and its applicability to university contexts. In the validation study, it also predicted intention to continue using institutional technology \cite{Boe2018}.

\paragraph{Hallucinations} We investigated users' cautious behavior toward LLM hallucinations, including their verification practices and task-specific trust decisions, with three items (\eg, ``I check important information from [the university chatbot/ChatGPT] responses through external sources.''). Users also reported their perceived hallucinations with three items (\eg, ``I have noticed that [the university chatbot/ChatGPT] makes up facts that do not correspond to reality.''). The items were self-developed, as no existing validated scale captured these aspects of LLM hallucinations.
Two of the three items measuring perceived hallucinations build on the first two items of the ``AI Hallucination Potential'' scale by \citet{Christensen2024}, which we adapted to our context. We added a third item based on the definition of hallucinations provided by \citet{Huang_2025}. 
The items measuring cautious behavior toward hallucinations capture three key user strategies: deciding on LLM use based on the importance of factual accuracy, verifying outputs externally, and adjusting prompts to improve reliability. These strategies are informed by existing research on LLM hallucination mitigation \cite{barkley2024, gordon2023}.
The internal consistency of the perceived hallucinations scale was acceptable, while cautious behavior toward hallucinations showed low internal consistency. Item-level analysis indicated that removing individual items did not increase reliability beyond $\alpha = 0.41$. We discuss this limitation in \autoref{Limitations}. Discriminant validity between the two constructs was supported by low correlations, indicating that they represent distinct constructs.

\paragraph{Privacy concerns} We measured privacy concerns with three items adapted from \citet{Hsu2018} (\eg, ``My decision to use [the university chatbot/ChatGPT] exposes me to privacy risks.''). To increase content validity and contextual relevance for our study, we added two self-developed items. One assessing users’ belief in the confidentiality of the system; the other concerns about chat conversations being read by third parties. This addition was motivated by recent research suggesting that such concerns may be particularly relevant for students in academic contexts \citep{Chen2023}.

\paragraph{Sustainability-aware AI usage} 
\mbox{}\\
We measured the concept of sustainability-aware AI usage with a three-item scale capturing environmentally conscious behavior and energy-related considerations (\eg, ``I often think about the energy consumption of [the university chatbot/ChatGPT].'').
As no validated instruments for the sustainable use of LLM chatbots currently exist, the items were self-developed. Content validity of the scale was established through systematic consideration of key dimensions of environmentally conscious AI interaction behavior based on literature review \cite{Almanasreh.2019}. The items capture three facets of sustainable AI usage: resource-efficient prompting (message length consideration), environmental impact awareness (energy consumption consciousness), and deliberate ecological decision-making (task-specific usage evaluation) \cite{Sanguinetti.2024, Graves.2025}. 

Since the customized LLMaaS displays the remaining number of tokens as a percentage (see \autoref{fig:User_Interface}, label 2), we additionally assessed the customized LLMaaS' users' awareness of token usage with two items (\eg, ``During use, I pay attention to how many tokens I still have available.'').

\subsection{Participants}
Participants were recruited via the university chatbot landing page, departmental communications, campus flyers, the university news website, and official university social media channels. A total of $N = 526$ participants took part in the survey. $n=116$ participants used both systems regularly. We determined our target sample size through a priori power analysis ($f^{2} = 0.05$, $\alpha = .05$, $\beta = .80$), which suggested N = 250 for within-subject analyses. The achieved subsample provides sufficient power to detect medium-sized effects, although smaller effects may remain undetected.
Within the total sample, 186 participants completed the questionnaire for the university's chatbot (customized LLMaaS) and 333 for ChatGPT (commercial LLM chatbot). An additional 123 participants reported using neither system at least once per month. \autoref{fig:Participants_by_System} shows an overview of participant groups according to their system usage.

Compared to the total sample, the subsample of dual-system users included fewer students and more research and administrative staff (see \autoref{tab:sample_demogra_table} for an overview and comparison of demographic characteristics). This subgroup also had a slightly higher proportion of male participants and was older on average. Within this subsample, $10.3\%$ reported using a paid ChatGPT plan, while $87.1\%$ used the free version (see \autoref{tab:chatgpt_plan}).

\subsection{Open Science}
\label{open_science}

This study was preregistered on OSF\footnote{\url{https://osf.io/kjcg5/overview?view_only=2e3c090e4d294fe8b714550a714bcb1e}}. Minor linguistic and structural adjustments were made to the preregistered hypotheses to improve clarity and consistency with the final study design. The dataset and supplementary materials (codebook, questionnaire overview) are available at OSF\footnote{\url{https://osf.io/vqsem/files/osfstorage?view_only=53ba0188a6db4a54aa08a1f34bf9078f}}. We excluded privacy-sensitive entries from the dataset prior to publication.

%% file: section/04_results.tex
\section{Results}

\subsection{Descriptive Results: Chatbot Usage Patterns and Knowledge Test}
Among the subsample of participants who had used both systems ($n = 116$), the university chatbot was most commonly used several times a month ($32.8\%$) or several times a week ($25.0\%$). In contrast, ChatGPT had a higher proportion of frequent users, with $19.0\%$ using it several times a day and $28.4\%$ several times a week (see \autoref{fig:Usage_Frequency_by_System} for full distribution).
Regarding usage purposes, both systems were most often used for reviewing or improving texts ($70.7\%$), text creation ($54.3\%$ university chatbot; $61.2\%$ ChatGPT) and information search ($52.6\%$ university chatbot; $67.2\%$ ChatGPT; see \autoref{fig:Purpose_of_Use_by_System}).
Awareness of model-switching functionality in the university chatbot was relatively high: $78.4\%$ of dual users knew about this option. However, only $41.4\%$ had actively switched from the default model (GPT-4o).
Regarding performance on the knowledge test about the university’s customized LLMaaS, $24.1\%$ answered four questions correctly and $21.6\%$ achieved a perfect score. $14.7\%$ did not answer any item correctly (see \autoref{tab:knowledge_score_distribution}).

\subsection{Hypothesis Testing: Comparisons of the Customized LLMaaS Chatbot vs. ChatGPT}

\subsubsection{Differences in Perceived Trust, Hallucinations, Privacy and Sustainability-Aware Use}
We conducted paired t-tests (\autoref{tab:ttest_comparisons}) to test H1 and H3-H6.
To account for the increased risk of Type I error due to multiple hypotheses testing, we applied Holm’s step-down correction to the family of six statistical tests \cite{BENDER2001, holm1979}. This procedure controls the family-wise error rate (FWER) at $\alpha = 0.05$ and was chosen because it is uniformly more powerful than the classical Bonferroni correction while still strictly controlling the FWER~\cite{BENDER2001}. The six $p$-values were ordered from smallest to largest and sequentially compared against adjusted α-thresholds (\ie, $\alpha$/$k$, where $k$ is the number of remaining tests). The procedure was terminated once a $p$-value exceeded the corresponding adjusted α-threshold; the respective hypothesis and all subsequent ones were rejected. 
H6, concerning sustainability-aware AI use, was the first hypothesis to exceed its adjusted α-threshold ($p = 0.03$ vs. $\alpha = 0.0167$), resulting in its rejection along with all subsequent hypotheses. 
Consequently, H3, concerning cautious behavior toward hallucinations ($p = 0.19$) and H2 (moderation: system type $\times$ trust in the university; $p = 0.44$) were also rejected. 

H1, H4 and H5 were supported, as the p-values remained below the respective adjusted $\alpha$-threshold: participants reported significantly (H1) higher levels of trust, (H4) fewer perceived hallucinations and (H5) fewer privacy concerns for the customized LLMaaS compared to ChatGPT.

\input{tables/t_test_table}

\subsubsection{Moderating Role of Trust in the University} 
\label{sec:moderation_trust_university}
To test whether trust in the university moderated the system type effect on user trust (H2), we fitted a Linear Mixed-Effects Model (see \autoref{tab:interaction_analysis_trust_in_university}) using the lme4 package \cite{bates2015lme4}. We used a Mixed-Effects Model rather than a simpler linear model (as preregistered) to properly account for the non-independence of the two trust ratings provided by each participant. The model predicted trust from system type (0~=~ChatGPT, 1~=~customized~LLMaaS~chatbot), mean-centered trust in the university (to improve interpretability of the intercept), and their interaction, with a random intercept for each participant to capture between-person variability.

\input{tables/model_orgatrust_mixed_centered_custom.tex}

H2 was not supported, as the interaction between system type and trust in the university was not significant. This indicates that the positive relationship between trust in the university and system trust did not differ between ChatGPT and the university’s customized LLMaaS.

\subsection{Controlling for Confounding Variables}
\label{ControlVariables}

We extended our preregistered analyses to account for potential confounding variables. In this section, we report robustness checks that address factors which could plausibly influence our main hypothesis tests, while the following section covers exploratory analyses aimed at uncovering additional patterns beyond our preregistered scope.

\subsubsection{Usage Frequency does not moderate the relationship between system type and perceived hallucinations}
\label{sec:usage_frequency_moderation}
To test whether the difference in perceived hallucinations between the customized LLMaaS chatbot and ChatGPT was moderated by usage frequency, we estimated a Cumulative Logit Model (\autoref{tab:vgam_hallucinations}) including both main effects and their interaction using the VGAM package in R \cite{yee2010vgam}. Usage frequency did not significantly moderate the relationship between system type and perceived hallucinations, as indicated by a Wald test of all interaction terms ($\chi^2 = 9.60$, $df = 5$, $p = 0.09$).
We chose this modeling approach because the usage frequency was measured with an ordinal scale. We accordingly transformed hallucination ratings into five ordered categories using equally spaced cut-off points.
Among the usage frequency contrasts, the lowest p-value was observed for the fifth-order polynomial, which tests whether hallucination reports follow a complex pattern across usage levels. However, this effect did not fall below the alpha level of $0.05$ and was therefore not statistically significant.

\input{tables/vgam_hallucinations_table.tex}  

\subsubsection{Model switching does not affect system differences}
To examine whether switching the default language model (GPT-4o) within the university's customized chatbot influenced the results, we repeated the within-subject comparisons using paired-samples t-tests on a reduced subsample of participants who reported never having changed the underlying model ($n = 68$; see \autoref{tab:ttest_non_switchers} for detailed test statistics). The pattern of results remained identical: the same differences between systems reached statistical significance (or not), and all observed effect directions were consistent with those in the full sample.

\subsubsection{Prior knowledge does not affect system differences}
To control for potential confounding effects of participants' knowledge about the university's customized LLMaaS chatbot, we included knowledge scores as a covariate (using Linear Mixed-Effects Models with the lme4 package, with fixed effects for system type, knowledge scores, and their interaction). The system differences proved robust to the prior knowledge of the participants (all $p > 0.11$; see \autoref{tab:knowledge_control}).

\subsection{Assessment of Common-Method Bias}

Because all key constructs were measured via self-report, we examined the risk of common-method bias. We focused on analyses that modeled associations between multiple predictors (\ie, the mixed-effects moderation model in \autoref{sec:moderation_trust_university} and the ordinal regression in \autoref{sec:usage_frequency_moderation}), as within-subject mean comparisons are largely unaffected because any method factor cancels out in the difference scores. Following the full-collinearity approach, we computed variance inflation factors (VIFs) from equivalent linear models including all fixed effects \cite{kock2015}. In the mixed-effects model, all predictors (system type, trust in the university, and their interaction) showed VIFs $\leq 2.0$ (see \autoref{tab:vif_mixed_orgatrust}). In the ordinal model, adjusted generalized VIFs $\bigl(\operatorname{GVIF}^{1/(2 \cdot df)}\bigr)$ \cite{fox1992} for system type, usage frequency, and their interaction were $\leq 2.67$ (see \autoref{tab:gvif_vgam_frequency}). These values are well below conservative thresholds (3.3; \cite{Cenfetelli2009, Petter2007}) and indicate no problematic multicollinearity, nor evidence of a dominant common-method factor.

\subsection{Exploratory Analyses}
\label{Exploratory Analyses}

\subsubsection{Benchmark evaluations reveal more hallucination generation but better hallucination recognition for the customized LLMaaS chatbot}
\label{Exploratory Analysis for Hallucinations}
To complement the self-reported perceived hallucinations of users, we conducted two exploratory benchmark evaluations to assess the hallucination tendencies of the chatbots: TruthfulQA and HaluEval. The university's customized LLMaaS chatbot and ChatGPT were tested using the same LLM, namely GPT-4o. 

\paragraph{TruthfulQA}
TruthfulQA \cite{lin-etal-2022-truthfulqa} is a benchmark designed to evaluate whether language models generate factually accurate responses to questions that are prone to false answers due to common misconceptions. A model performs well if it avoids reproducing widespread falsehoods. We evaluated both systems on 50 questions from the TruthfulQA dataset including 19 questions from the ``Education \& Science'' category to reflect the university context of our study and 31 questions from the remaining categories.
The university's customized LLMaaS chatbot produced hallucinated answers in 25 out of 50 cases ($50\%$), whereas ChatGPT hallucinated in 19 out of 50 cases ($38\%$). Notably, the university chatbot hallucinated more frequently across both the domain-specific and general subsets of the benchmark.

\paragraph{HaluEval}
HaluEval \cite{li-etal-2023-halueval} is a large-scale collection of generated and human-annotated hallucinated samples for evaluating the performance of LLMs in recognizing hallucinations. 
We tested both systems on $10,000$ QA examples from the HaluEval dataset. Results showed that both systems achieved comparable performance, with the university's LLMaaS chatbot reaching slightly higher values than ChatGPT across all metrics. 
Specifically, it achieved higher accuracy ($0.7190$ vs. $0.7011$), precision ($0.6671$ vs. $0.6564$), recall ($0.8807$ vs. $0.8472$), and F1-score ($0.7591$ vs. $0.7397$).

\subsubsection{Perceived usefulness does not differ, ease of use is greater for the customized LLMaaS chatbot} 
For participants who had used both systems, paired t-tests were conducted to compare perceived usefulness and ease of use (see \autoref{tab:ttest_comparisons}).
For usefulness, no statistically significant difference was found between the customized LLMaaS ($M = 4.01$, $SD = 0.71$) and ChatGPT ($M = 3.98$, $SD = 0.82$), $t(115) = 0.41$, $p = 0.68$, $d = 0.04$.
For ease of use, ratings were statistically significantly higher for the customized LLMaaS ($M = 4.59$, $SD = 0.65$) compared to ChatGPT ($M = 4.48$, $SD = 0.64$), $t(115) = 2.12$, $p = 0.04$, $d = 0.20$.

\subsubsection{Between-subjects comparisons mostly align with within-subjects results} 
To complement the within-subject analyses, we conducted exploratory between-subject comparisons using Welch’s two-sample t-tests. These analyses included participants who reported regular use of only one system -- either the university's customized LLMaaS chatbot ($n = 70$) or ChatGPT ($n = 217$). The full set of results is provided in \autoref{tab:ttest_between_table}; overall, the between-subjects results mostly aligned with the within-subjects results, and here we report only findings that deviated. 
Participants who exclusively used the customized LLMaaS chatbot reported significantly higher sustainability-aware AI usage compared to those who only used ChatGPT. The significant within-subject difference in trust in favor of the customized LLMaaS chatbot was not replicated. Similarly, the significant within-subject effect for perceived ease of use (favoring the customized LLMaaS chatbot) was not statistically significant.
Given that ChatGPT Pro users are not subject to token limits, which may compromise the comparability of groups for sustainability-aware AI usage, we conducted an additional comparison between exclusive ChatGPT Free users ($n = 176$) and users of the customized LLMaaS chatbot ($n = 70$). The same pattern of results were obtained as with the full-sample (see \autoref{tab:ttest_between_free}).

\subsubsection{Customized LLMaaS chatbot is used mainly for academic purposes, ChatGPT for both academic and private purposes} 
\label{usage_context_results}

Participants were instructed to answer items with regard to their study- or work-related use of the AI systems.
However, we also asked about the overall usage tendencies.
Participants who used both systems reported using the customized LLMaaS chatbot with a strong orientation toward study- or work-related contexts ($M = 5.48$, $SD = 0.81$; scale range: $1-6$), whereas reported usage of ChatGPT was more balanced across the spectrum ($M = 3.32$, $SD = 1.61$). A paired samples t-test confirmed that this difference was statistically significant, $t(115) = 13.36$, $p < .001$, $d = 1.24$.

%% file: tables/t_test_table.tex

\begin{table*}[t]
\centering
\begin{minipage}{\textwidth} 
\centering
\caption{\textbf{Paired t-test comparisons of the customized LLMaaS chatbot (UniGPT) vs. ChatGPT (n = 116), ordered by p-values used for Holm correction.}}
\label{tab:ttest_comparisons}
\small
\setlength{\tabcolsep}{2.5pt}

\begin{tabular*}{\textwidth}{@{\extracolsep{\fill}}lcccccccccc}
\toprule
\textbf{Dependent Variable} & \textbf{\makecell{\textit{M}\\(UniGPT)}} & \textbf{\makecell{\textit{SD}\\(UniGPT)}} & \textbf{\makecell{\textit{M}\\(ChatGPT)}} & \textbf{\makecell{\textit{SD}\\(ChatGPT)}} & \textbf{\textit{t}} & \textbf{\textit{df}} & \textbf{\textit{p}} & \textbf{Holm $\alpha$\textsuperscript{1}} & \textbf{\textit{d}\textsuperscript{2}} & \textbf{95\% \textit{CI}\textsuperscript{3}} \\
\midrule
Privacy concerns (H5) & 2.38 & 0.95 & 3.61 & 0.85 & -12.42 & 115 & < 0.001 & 0.0083 & -1.15 & [-1.39, -0.92] \\
Trust (H1) & 3.60 & 0.72 & 3.08 & 0.79 & 7.84 & 115 & < 0.001 & 0.01 & 0.73 & [0.52, 0.93] \\
Perceived hallucinations (H4) & 3.23 & 1.00 & 3.85 & 0.85 & -7.76 & 115 & < 0.001 & 0.0125 & -0.72 & [-0.92, -0.52] \\
Sustainability-aware AI use (H6) & 2.33 & 1.18 & 2.22 & 1.17 & 2.16 & 115 & 0.03 & 0.0167 & 0.20 & [0.02, 0.38] \\
Cautious behavior toward hallucinations (H3) & 4.18 & 0.68 & 4.13 & 0.67 & 1.31 & 115 & 0.19 & n/a\textsuperscript{4} & 0.12 & [-0.06, 0.30] \\
\midrule
Usefulness\textsuperscript{5} & 4.01 & 0.71 & 3.98 & 0.82 & 0.41 & 115 & 0.68 &  & 0.04 & [-0.14, 0.22] \\
Ease of use\textsuperscript{5} & 4.59 & 0.65 & 4.48 & 0.64 & 2.12 & 115 & 0.04 &  & 0.20 & [0.01, 0.38] \\
\bottomrule
\end{tabular*}

\vspace{2pt}
\footnotesize
\raggedright
\textit{Notes.}
\textsuperscript{1} Holm-adjusted significance threshold per hypothesis rank (sequential correction).
\textsuperscript{2} Cohen’s \textit{d} is reported as the standardized effect size for each comparison.
\textsuperscript{3} Confidence intervals (CI) for Cohen's d are not adjusted for multiple comparisons.
\textsuperscript{4} Holm-adjusted thresholds are only assigned while p-values remain below $\alpha$; correction stops after the first non-significant test.
\textsuperscript{5} Ease of Use and Usefulness were analyzed exploratively and were not part of the Holm correction.
\par
\end{minipage}
\end{table*}

%% file: tables/model_orgatrust_mixed_centered_custom.tex
\begin{table}[htbp]
\centering
\caption{\textbf{Mixed-Effects Model of system and trust in the university on user trust (number of observations: 232; $\textbf{n} = \textbf{116}$).}}
\label{tab:interaction_analysis_trust_in_university}
\small
\setlength{\tabcolsep}{2pt}
\begin{tabular}{l D{.}{.}{3.3} D{.}{.}{3.3} D{.}{.}{2.1} D{.}{.}{2.3} D{.}{.}{3.3}}
\toprule
\textbf{Effect} & \multicolumn{1}{c}{\textit{Estimate}} & 
\multicolumn{1}{c}{\textit{SE}} & 
\multicolumn{1}{c}{\textit{df}} & 
\multicolumn{1}{c}{\textit{t}} & 
\multicolumn{1}{c}{\textit{p}} \\
\midrule
(Intercept) & 3.075\textsuperscript{***} & 0.066 & 184.7 & 46.33 & <0.001 \\
System & 0.528\textsuperscript{***} & 0.067 & 114.0 & 7.83 & <0.001 \\
TrustUniversity\_centered & 0.272\textsuperscript{**} & 0.084 & 184.7 & 3.25 & 0.001 \\
System:TrustUniversity\_centered & 0.066 & 0.085 & 114.0 & 0.78 & 0.440 \\
\midrule
\multicolumn{6}{l}{\textbf{Model fit}} \\
AIC & \multicolumn{5}{c}{493.72} \\
BIC & \multicolumn{5}{c}{514.40} \\
Log Likelihood & \multicolumn{5}{c}{-240.86} \\
\midrule
\multicolumn{6}{l}{\textbf{Random effects}} \\
Num. groups: Bogen & \multicolumn{5}{c}{116} \\
Var: Bogen (Intercept) & \multicolumn{5}{c}{0.25} \\
Var: Residual & \multicolumn{5}{c}{0.26} \\
\bottomrule
\multicolumn{6}{l}{\footnotesize{$^{***}p<0.001$; $^{**}p<0.01$; $^{*}p<0.05$}} \\
\end{tabular}
\end{table}

%% file: tables/vgam_hallucinations_table.tex
\begin{table}[htbp]
\caption{\textbf{Cumulative Logit Model: System type and usage frequency on perceived hallucinations (number of observations: 227).}}
\label{tab:vgam_hallucinations}
\small

\begin{minipage}{\columnwidth}
\centering

\begin{tabular}{l D{.}{.}{3.3} D{.}{.}{3.3} D{.}{.}{2.2} D{.}{.}{3.3}}
\toprule
\textbf{Effect} & \multicolumn{1}{c}{\textit{Estimate}} & 
\multicolumn{1}{c}{\textit{SE}} & 
\multicolumn{1}{c}{\textit{z}} & 
\multicolumn{1}{c}{\textit{p}} \\
\midrule
System (ChatGPT vs. UniGPT) & -0.955\textsuperscript{**} & 0.309 & -3.09 & 0.002 \\
Frequency (Linear) & -0.613 & 0.539 & -1.14 & 0.255 \\
Frequency (Quadratic) & 0.004 & 0.472 & 0.01 & 0.994 \\
Frequency (Cubic) & -0.511 & 0.556 & -0.92 & 0.358 \\
Frequency (4\textsuperscript{th} order) & -0.095 & 0.555 & -0.17 & 0.864 \\
Frequency (5\textsuperscript{th} order) & -0.811 & 0.452 & -1.79 & 0.073 \\
System × Frequency (Linear) & 0.954 & 0.743 & 1.28 & 0.199 \\
System × Frequency (Quadratic) & -0.912 & 0.653 & -1.40 & 0.163 \\
System × Frequency (Cubic) & -0.184 & 0.809 & -0.23 & 0.821 \\
System × Frequency (4\textsuperscript{th} order) & -1.158 & 0.832 & -1.39 & 0.164 \\
System × Frequency (5\textsuperscript{th} order) & 0.171 & 0.683 & 0.25 & 0.802 \\
\midrule
\multicolumn{5}{l}{\textbf{Model fit}} \\
Residual Deviance & \multicolumn{4}{c}{608.20} \\
Log Likelihood & \multicolumn{4}{c}{-304.10} \\
AIC & \multicolumn{4}{c}{638.20} \\
\bottomrule
\end{tabular}

\vspace{2pt}
\footnotesize
\raggedright
\textit{Notes.} $^{***}p<0.001$; $^{**}p<0.01$; $^{*}p<0.05$. \\
Model estimated on 227 observations (df = 893).
\par

\end{minipage}
\end{table}

%% file: section/05_discussion.tex
\section{Discussion}

This study examined how customized LLMaaS chatbots are perceived and used differently compared to commercially available alternatives in the context of everyday university life. Our findings support the hypothesized differences in users' perceptions in regard to trust, privacy and perceived hallucinations.

\subsection{Understanding User Perceptions of a Customized LLMaaS Chatbot vs. ChatGPT} 
The findings demonstrate that students and staff perceive the customized LLMaaS chatbot differently compared to ChatGPT. This is a significant finding, as educational institutions are increasingly making high investments in generative AI solutions \cite{uspenskyi2025main}, and there has been no comparative study of the perception of such solutions from the user's perspective to date.

\paragraph{Participants showed higher trust (H1) and fewer privacy concerns (H5) with regards to the university's customized LLMaaS chatbot compared to ChatGPT} It is conceivable that the differences between the systems enhance user trust and perceived privacy by increasing factors such as perceived openness or transparency of the system, which is consistent with previous findings \cite{schwartz2023, nordheim2019}. Additionally, users might have associated the university-hosted system with more responsible or secure data practices due to its institutional branding in the interface design or login-based access within the university network. This adds to general work from marketing research that showed that strong corporate branding can reduce perceived privacy risks and increase users’ willingness to disclose personal information online~\cite{wang2019, myerscough2008, jain2022}.

\paragraph{The hypothesized moderation effect of trust in the university (H2) was not supported.} The effect of system type on user trust did not depend on how much users trusted the university. Instead, both the system type (customized LLMaaS chatbot) and higher trust in the university independently contributed to higher trust in the system. This finding may reflect the influence of users’ general propensity to trust, that is, a dispositional tendency to attribute trustworthiness across different social and technological entities \cite{McKnight2011, Heyns}. Overall, the results point to additive rather than interactive effects of system type and trust in the university on users' trust in the system.

\paragraph{Participants did not behave more cautiously toward hallucinations (H3), but reported more perceived hallucinations with ChatGPT than with the customized LLMaaS chatbot (H4).}
This combination of results is noteworthy as we expected that the LLMaaS chatbot would jointly be associated with less cautious behavior and fewer perceived hallucinations compared to ChatGPT, both resulting from lower user alertness toward hallucinations. However, our findings only support fewer perceived hallucinations (H4). While the lack of an effect for H3 could be due to the low internal consistency of our self-developed scale, there is another possible explanation: The interface customizations may have reduced users' mental alertness only to a limited degree, influencing the detection of hallucinations, which rely more on automatic or intuitive processes, but not more reflective, effortful strategies for preventing hallucinations or verifying outputs.

Concentrating on H4, this finding is particularly interesting when considered alongside our exploratory benchmark evaluations. TruthfulQA directly assessed hallucination generation and indicated that the LLMaaS chatbot actually produced more hallucinated answers than ChatGPT. Because the customized LLMaaS chatbot operated at a lower temperature than ChatGPT, one might expect it to hallucinate less. Yet, this benchmark suggests that the temperature difference did not reduce hallucination generation and therefore cannot account for the lower level of perceived hallucinations reported by participants. HaluEval evaluated hallucination detection performance and showed a minor difference, with the university chatbot outperforming ChatGPT. Taken together, these results suggest a miscalibration between users’ perceptions and objective performance: participants apparently notice less hallucinations in the customized LLMaaS chatbot, although benchmark evidence indicates that it may generate hallucinations more frequently. This discrepancy may indicate that the institutional framing and the absence of an in-chat warning could have reduced users’ attentiveness to hallucinations, thereby contributing to an underestimation of risks.

\paragraph{Participants did not report more sustainability-aware AI usage when interacting with the customized LLMaaS chatbot compared to ChatGPT (H6).} This finding stands in contrast to previous work showing that making resource consumption visible can foster more sustainable behavior \cite{Penkert.2023, Sanduleac.2017}. The exploratory between-subjects comparisons showed that users who exclusively used the customized LLMaaS chatbot reported more sustainability-aware AI usage than those who exclusively used ChatGPT. This could reflect underlying differences between user groups, e.g., regarding their values \cite{Shang2023} or usage motivations \cite{Wunderlich2019}. Alternatively, lower sustainability-aware AI usage associated with ChatGPT might originate from characteristics of the system itself (\eg, no token usage visualization; low-barrier access across devices). The results highlight a promising avenue for future research on how individual traits or system characteristics may shape sustainability-aware AI usage.

\subsection{Implications}
\subsubsection{Theoretical Implications}

We draw on the TrAM model \cite{schlicker2025trustworthiness} to theorize why there might be differences in perceptions between customized LLMaaS and commercially available alternatives. Specifically, we theorize that perceived system characteristics only partially link to actual system characteristics and that there are cues experienced by users that lead to differences in perceived system characteristics that are unrelated to the actual system characteristics. We refer to the concept of calibration to describe the resulting fit between actual system characteristics and perceived system characteristics, a concept that has previously been considered primarily in relation to trust \cite{ParasuramanRiley1997, lee2004trust}. 

In general, we see that users displayed higher trust, perceived greater privacy and perceived less hallucinations with the customized LLMaaS chatbot. At first glance, this suggests that users are able to calibrate their perceptions towards an LLM chatbot, as, \eg, the university's customized chatbot has stronger privacy safeguards and users have less concerns. 
However, we question whether the extent of the perceived difference is justified and if users' consequent behavior is ``right''. First, we question how well users understand the differences between the systems, given the results of the knowledge test (see \autoref{ControlVariables}). Second, an appropriate calibrated perception is further questioned by the results of the hallucination benchmark evaluations, which allowed us to relate users’ subjective experiences to objective performance metrics. Participants perceived fewer hallucinations with the customized LLMaaS, although the TruthfulQA benchmark suggests that this system may actually be more prone to hallucinations. Accordingly, the anticipated link between the system’s lower temperature setting and reduced hallucination tendency (see \autoref{fig:TrAM_Figure}) did not hold. In sum, this divergence highlights a critical gap in users' hallucination alertness calibration, that is, their ability to align vigilance against hallucinations with the system’s actual risk. Users may underestimate the risk of hallucinations due to trust-inducing cues such as branding, institutional context, and lack of persistent warnings. When this finding is considered alongside the significantly higher perceived privacy in LLMaaS, the question arises as to whether users are overtrusting the customized LLMaaS chatbot.

Taken together, the findings suggest that the TrAM is useful for identifying points in the user experience where misalignments between users’ perceptions and the system’s actual characteristics may occur. This makes it a valuable framework for scholars and practitioners to use when identifying or refining existing cues, or when introducing new ones, in order to foster more calibrated user perceptions.

\paragraph{Toward a Causal Understanding of User Perception Differences.} Our findings also open up avenues for future research on the cues underlying user judgments of AI systems. While we showed that user perceptions differed and we offer theoretically grounded interpretations as to how system customizations might have contributed to the results, the causality of the findings, e.g., with regards to the question of which specific cues users rely on most when forming such judgments, remains an open question. A key next step is to experimentally and systematically compare customization cues in terms of their influence on user perceptions and their alignment with actual system properties to better understand how (mis)calibrations emerge in practice. Future studies might incorporate experimental lab settings to test mediation analysis to model the causal relations between system type as experimental conditions, customization choices as mediators and user perceptions as outcomes.

\paragraph{New Directions for User Hallucination Detection Research}
This work stimulates new directions for research on users' detection of LLM hallucinations, particularly regarding usage context and task characteristics. Prior work has mainly focused on individual-level factors \cite{dang2025}, explanation formats \cite{Reinhard2025}, and information characteristics \cite{Tao2024}, but has paid less attention to how system use itself shapes hallucination detection. Our exploratory results point to notable differences in the broader usage context of the two systems (see \autoref{usage_context_results}). It remains an open question whether such contextual differences affect users’ expectations, vigilance, or standards for verifying outputs. Future research should also examine how finer-grained usage purposes -- such as task complexity, topical domain, or goal type (\eg, idea generation vs. factual retrieval) -- shape users’ scrutiny of outputs, providing a more comprehensive understanding of hallucination detection.

\subsubsection{Practical Implications}

Our study provides valuable insights for organizations considering the adaption of a customized LLMaaS chatbot and those actively running one.

We advise institutions deploying LLMaaS chatbots to pay attention to the potential discrepancy between the actual and perceived trustworthiness of the system. In particular, developers and deployers should ensure that surface-level cues do not create a false impression of the system's capabilities or privacy safeguards. This could involve actively communicating residual risks, providing context-specific warnings, and offering transparency tools. 
The results of our study showed that although a system might not hallucinate less, it can be perceived to do so. Therefore, we suggest implementing a hallucination warning in the user interface. Previous research has shown that warnings can increase hallucination detection without significantly affecting perceived truthfulness of genuine content \cite{nahar2024}. Similarly, \citet{Bo2025} found that adding a persistent disclaimer was the most effective at improving users’ appropriate reliance in a complex problem-solving task. Building on this, additional warnings could be provided when the system offers links, encouraging users to verify their accuracy and relevance. Moreover, confidence scores and links connecting system responses to the source documents used to generate them could be effective transparency tools for supporting users' trustworthiness assessment process ~\cite{Leiser2023, Reinhard2025}. 

At an institutional level, measures could include a mandatory online course providing information on the risks, limitations, and particularities of the system, which users would have to complete to activate their account for using LLMaaS chatbots.
One helpful customization to support users' trustworthiness assessment process would be to integrate a dedicated “audit-assist” mode with structured, example-driven scaffolding features that guide users through systematic audits of model behavior. This can improve students’ critical engagement with LLMs \cite{Prabhudesai2025}.
Furthermore, alerts could be added to the interface when users ask for medical, legal or financial advice, or when they enter personal data, to facilitate output evaluation in high-risk contexts. 
To encourage appropriate user behavior particularly in relation to privacy, one option for developers of LLMaaS chatbots would be to integrate user-led data minimization tools that help users detect and remove personal information from their prompts~\cite{Zhou_Xu_Wu_Li_2025}.

\subsection{Limitations} 
\label{Limitations}

First, our main analyses were based on a within-subjects subsample of users who had experience with both systems. While this enables controlled comparisons, it may also introduce a self-selection bias. Although key effects (\eg, privacy and hallucinations) were replicated in the between-subjects analysis, sustainability-aware AI usage was only significant in the between-subjects comparison. Future studies should therefore examine the generalizability of our findings across user groups (\eg, students vs. employees; single-system vs. multi-system users). 
The generalizability of the results must also be discussed with regards to other (a) universities, (b) professional contexts, and (c) nationalities. In principle, case studies are transferable to the same extent as large-scale studies, when taking into account how similar the socio-organizational structures in another setting are \cite{tsang2014generalizing}. As many German universities currently introduce LLMaaS (without fine-tuning models), the results of this study are likely representative, if similar customization choices are made. It is also conceivable that similar effects are observable in other knowledge-work intense professions. However, results might not transfer to other domains e.g., healthcare, where for example reliability and accuracy of systems are large concerns \cite{chen2025systematic}. Additionally, Germany is characterized by strict data protection rules. Universities/ organization need to act in accordance with the General Data Protection Regulation (GDPR), which students and staff are likely aware of. This limits generalization to other (especially non-European) countries, particularly with regard to the perceived privacy of systems.

Second, we did not control for potential order effects in the questionnaire. Due to the limited configuration options on the university's internal survey platform, the university's LLMaaS chatbot was always evaluated before ChatGPT. While this ordering was consistent for all participants and unlikely to explain system-specific effects alone, it remains a potential confound. 

Third, our study relied exclusively on self-report measures and exploratory benchmark evaluations. Future studies should incorporate behavioral data, such as prompt input, output verification actions, or system usage logs, to complement self-reports. Moreover, perceptual measures alone may not fully capture how customization relates to actual model behavior. Although we included two exploratory objective benchmark evaluations to assess hallucination tendencies and detection performance (see Section~\ref{Exploratory Analysis for Hallucinations}), these analyses were not part of the primary inferential design. Combining self-report measures with objective task-based performance metrics would help distinguish between effects on user perception driven by interface cues and those reflecting underlying system performance.

Furthermore, the internal consistency of the scale measuring cautious behaviour towards hallucinations was low. This may be due to heterogeneity in the construct, as it includes both proactive (\eg, adjusting prompts) and reactive (\eg, checking outputs) strategies. Future research should consider splitting and extending this scale to better to distinguish dimensions and improve psychometric quality. 

Fourth, our within-subjects sample (n = 116) fell short of the planned size (n = 250). While sufficient for detecting medium effects, smaller effects, such as differences in cautious behavior or sustainability-aware AI use, may have gone undetected.

Fifth, the study design does not allow for causal interpretations. While one key strength of our study is that we surveyed real users in a naturalistic setting, thereby generating findings with high ecological validity, this comes at a cost of lower internal validity. This could be addressed in future research through the use of controlled experimental designs isolating effects of individual features, while longitudinal studies may shed light on how user perceptions evolve over time. Such designs could also examine causal links between specific user perceptions, \eg, user trust and perceived hallucinations.

Sixth, the systems differed not only in branding and context, but also in available features, \eg, ChatGPT supports voice input, web search, and file uploads, which the university chatbot does not. While we did not focus on these differences, they may have influenced user perceptions such as ease of use or perceived privacy. Future research should examine how specific features shape system evaluations.

\subsection{Conclusion} 

We examined how customized LLMaaS chatbots are perceived and used differently compared to commercially available alternatives. Participants reported greater trust, fewer perceived hallucinations, and lower privacy concerns when using the university’s customized chatbot compared to ChatGPT -- despite both systems being based on the same underlying LLM technology. These differences hint at the potential influence of contextual and interface-level cues on users’ system assessments.
Our findings emphasize the importance of calibrated user perceptions not only with regard to trustworthiness, but also in relation to privacy and hallucination risk. In doing so, our work contributes to the TrAM and the concept of calibrated trust, expanding both to include more facets of human-AI interaction. 
Practically, we provide evidence-based guidance on AI implementation. The careful design of interfaces, features, and integration strategies for LLMaaS chatbots can promote their appropriate use in line with their risks and limitations.

We encourage future research to build on the results and to explore which specific cues users rely on when forming their perceptions and how such alignments affect downstream behavior in experimental settings to complement the findings of this work with causal data. Our work, therefore, opens new research directions to investigate how customization aspects impact users.

%% file: acknowledgements.tex
\begin{acks}
We thank Markus Pauly, Nico Föge, and Daniel Klippert for their support with the statistical analyses. 
We are grateful to Lin Kyi, Gabriel Lima, and Yixin Zou for their thoughtful feedback on the manuscript. 
We also acknowledge the support of the involved parties at Ruhr University Bochum, including the work councils, the data protection officer, the communication department, and the IT department. 
Finally, we thank all participants for taking part in the study.

This work was supported by the \grantsponsor{rctrust}{Research Center Trustworthy Data Science and Security}{https://rc-trust.ai} (\url{https://rc-trust.ai}), one of the Research Alliance Centers within the UA Ruhr (\url{https://uaruhr.de}).
\end{acks}

%% file: section/07_Appendix.tex
\clearpage
\onecolumn
\appendix
\section*{Appendix}

\renewcommand{\thetable}{A\arabic{table}}
\renewcommand{\thefigure}{A\arabic{figure}}
\setcounter{table}{0}
\setcounter{figure}{0}

\let\oldfigure\figure
\let\endoldfigure\endfigure
\renewenvironment{figure}{\begin{figure*}}{\end{figure*}}

\let\oldtable\table
\let\endoldtable\endtable
\renewenvironment{table}{\begin{table*}}{\end{table*}}

\section{Demographics}
\label{app:demographics}

\begin{figure}[htb]
\centering
\includegraphics[width=0.9\textwidth]{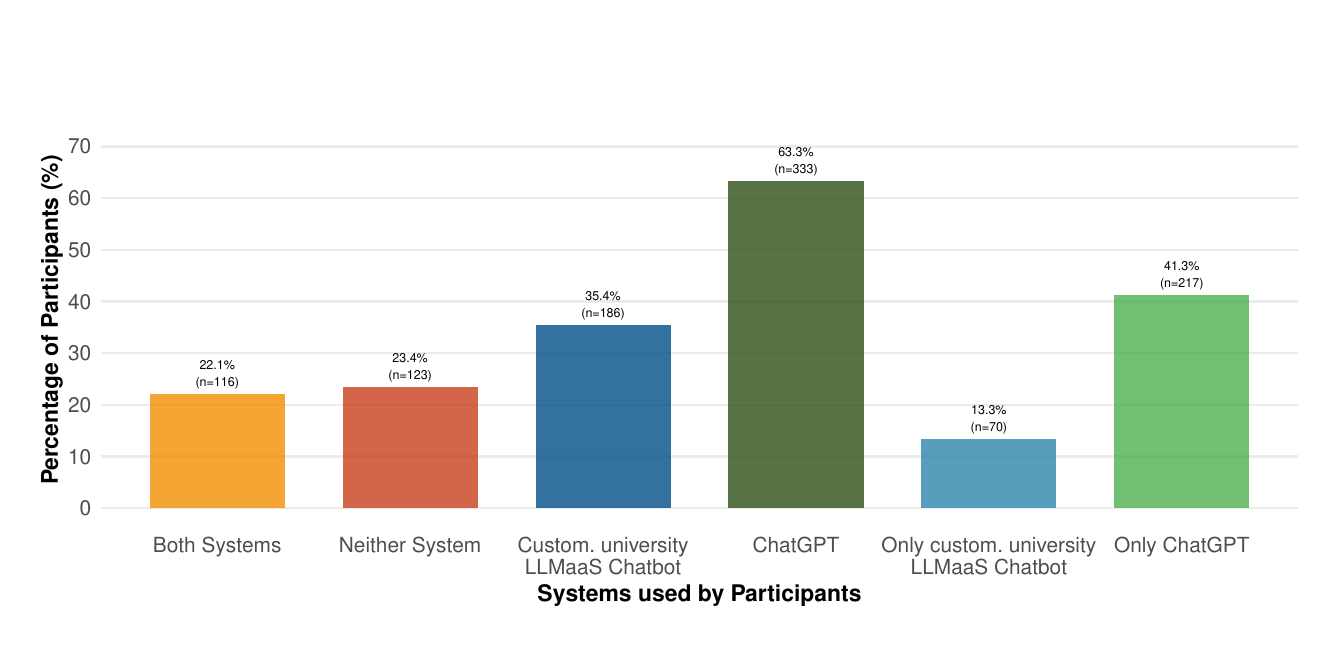}
\caption{\textbf{Percentage of participants using different chatbot systems.}}
\Description{Bar chart showing the percentage of participants for each usage category of the university’s customized LLMaaS chatbot and ChatGPT. 
The y-axis ranges from 0\% to 70\% in increments of 10\%. 
From left to right on the x-axis: 
Both Systems – 22.1\% (n = 116), 
Neither System – 23.4\% (n = 123), 
Custom university LLMaaS Chatbot – 35.4\% (n = 186), 
ChatGPT – 63.3\% (n = 333), 
Only custom university LLMaaS Chatbot – 13.3\% (n = 70), 
Only ChatGPT – 41.3\% (n = 217). 
Each bar is colored differently for visual distinction.}
\label{fig:Participants_by_System}
\end{figure}

\input{tables/Sample_Description_Table}

\input{tables/table_chatGPT_plan}

\newpage
\section{Correlation Tables}
\renewcommand{\thetable}{B\arabic{table}}
\renewcommand{\thefigure}{B\arabic{figure}}
\setcounter{table}{0}
\setcounter{figure}{0}

\input{tables/Correlations_CustomizedLLMaaS.tex}

\input{tables/Correlations_chatGPT.tex}

\newpage
\section{Descriptive Results}
\label{app:furtherdescriptiveresuts}
\renewcommand{\thetable}{C\arabic{table}}
\renewcommand{\thefigure}{C\arabic{figure}}
\setcounter{table}{0}
\setcounter{figure}{0}


\begin{figure}[htb]
\centering
\includegraphics[width=0.9\textwidth]{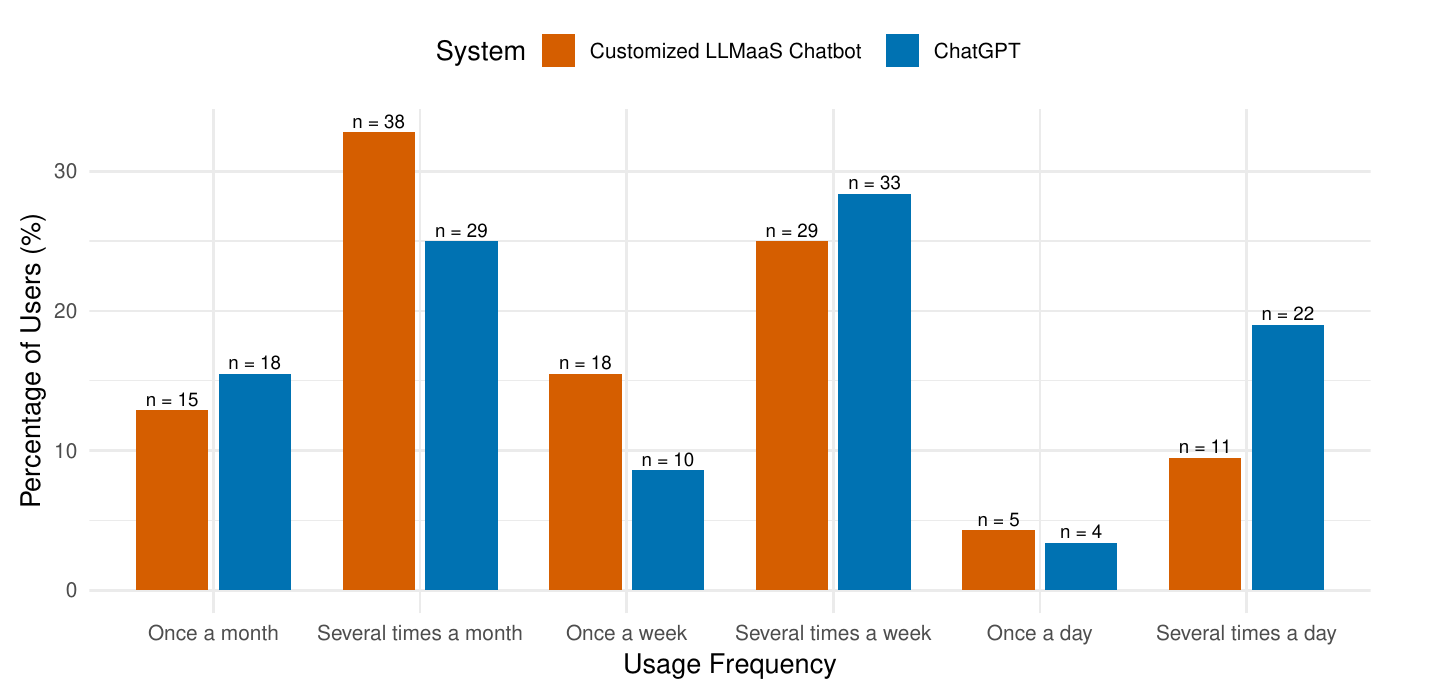}
\caption{\textbf{Usage frequency of the university’s customized LLMaaS chatbot vs. ChatGPT in the subsample of both systems users ($n = 116$).}}
\Description{Grouped bar chart comparing usage frequency for the university’s customized LLMaaS chatbot and ChatGPT. 
The customized chatbot is represented by orange bars, and ChatGPT by blue bars. 
The y-axis is labeled “Percentage of Users (\%)” and ranges from 0 to about 35 in regular increments. 
From left to right, the x-axis categories are: Once a month; Several times a month; Once a week; Several times a week; Once a day; Several times a day. 
Each bar is labeled with the count of users (n). 
Counts by category are:
Once a month — customized chatbot n=15, ChatGPT n=18;
Several times a month — customized chatbot n=38, ChatGPT n=29;
Once a week — customized chatbot n=18, ChatGPT n=10;
Several times a week — customized chatbot n=29, ChatGPT n=33;
Once a day — customized chatbot n=5, ChatGPT n=4;
Several times a day — customized chatbot n=11, ChatGPT n=22.}
\label{fig:Usage_Frequency_by_System}
\end{figure}

\begin{figure}[htb]
\centering
\includegraphics[width=0.9\textwidth]{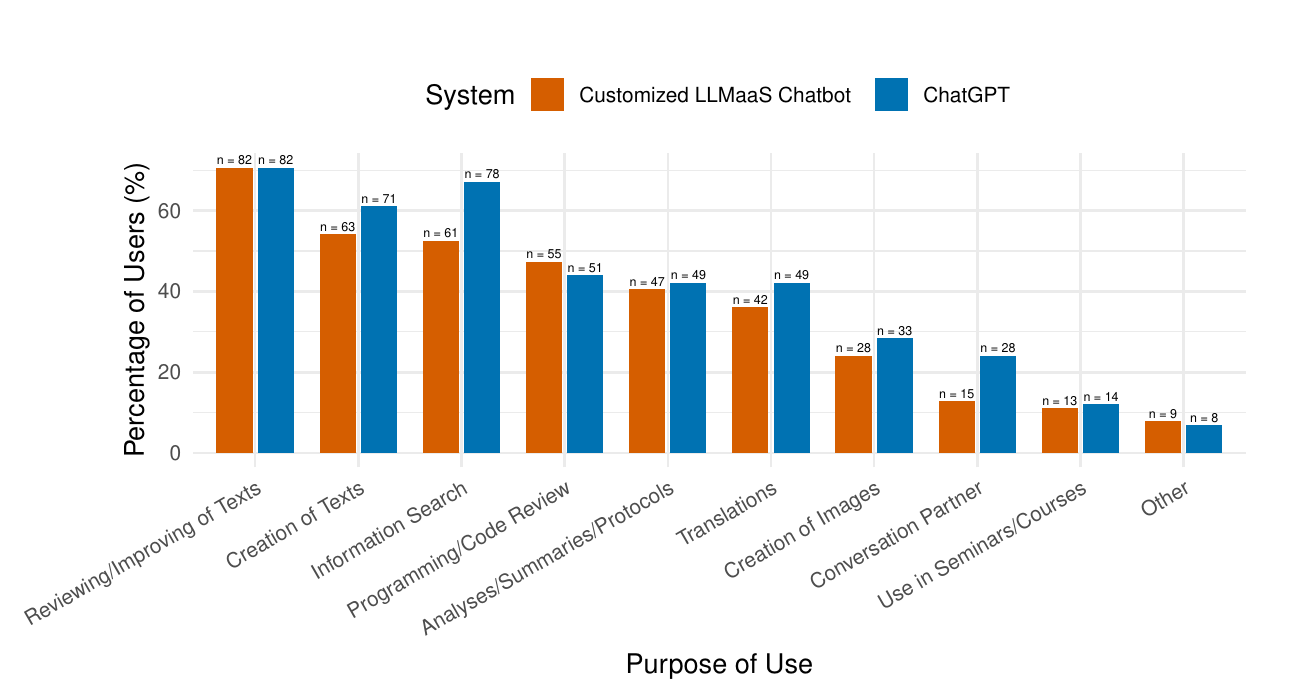}
\caption{\textbf{Purpose of use for the university’s customized LLMaaS chatbot vs. ChatGPT in the subsample of both systems users ($n = 116$).}}
\Description{Grouped bar chart comparing purposes of use for the university’s customized LLMaaS chatbot and ChatGPT. 
The customized chatbot is represented by orange bars, and ChatGPT by blue bars. 
The y-axis is labeled “Percentage of Users (\%)” and ranges from 0 to about 70 in increments of 10. 
From left to right, the x-axis categories are: Reviewing/Improving of Texts — customized chatbot n=82, ChatGPT n=82; 
Creation of Texts — customized chatbot n=63, ChatGPT n=71; 
Information Search — customized chatbot n=61, ChatGPT n=78; 
Programming/Code Review — customized chatbot n=55, ChatGPT n=51; 
Analyses/Summaries/Protocols — customized chatbot n=47, ChatGPT n=49; 
Translations — customized chatbot n=42, ChatGPT n=49; 
Creation of Images — customized chatbot n=28, ChatGPT n=33; 
Conversation Partner — customized chatbot n=15, ChatGPT n=28; 
Use in Seminars/Courses — customized chatbot n=13, ChatGPT n=14; 
Other — customized chatbot n=9, ChatGPT n=8.}
\label{fig:Purpose_of_Use_by_System}
\end{figure}

\input{tables/Knowledge_Scores_Distribution.tex}

\newpage
\section{Controlling for Confounding Variables}
\renewcommand{\thetable}{D\arabic{table}}
\renewcommand{\thefigure}{D\arabic{figure}}
\setcounter{table}{0}
\setcounter{figure}{0}

\input{tables/t_tests_no_switchers.tex}

\input{tables/knowledge_as_control_variable.tex}

\newpage
\section{Full-Collinearity Checks}
\renewcommand{\thetable}{E\arabic{table}}
\renewcommand{\thefigure}{E\arabic{figure}}
\setcounter{table}{0}
\setcounter{figure}{0}

\begin{table}[h]
\centering
\small
\caption{\textbf{Full-collinearity VIFs for the mixed-effects model (System × Trust in University).}}
\label{tab:vif_mixed_orgatrust}
\begin{tabular}{lc}
\toprule
\textbf{Predictor} & \textbf{VIF} \\
\midrule
System & 1.00 \\
Trust in University (centered) & 2.00 \\
System × Trust in University & 2.00 \\
\midrule
\textbf{Max VIF} & \textbf{2.00} \\
\bottomrule
\end{tabular}
\end{table}

\begin{table}[h]
\centering
\small
\caption{\textbf{Adjusted GVIFs for the ordinal regression model (System × Usage Frequency).}}
\label{tab:gvif_vgam_frequency}
\begin{tabular}{lccc}
\toprule
\textbf{Predictor} & \textbf{GVIF} & \textbf{Df} & \textbf{GVIF$^{1/(2 \cdot Df)}$} \\
\midrule
System & 7.12 & 1 & 2.67 \\
Usage Frequency (polynomial contrasts) & 1.21 & 5 & 1.02 \\
System × Usage Frequency & 8.31 & 5 & 1.24 \\
\midrule
\textbf{Max adjusted GVIF} & — & — & \textbf{2.67} \\
\bottomrule
\end{tabular}
\end{table}

\newpage
\section{Exploratory Analyses}
\renewcommand{\thetable}{F\arabic{table}}
\renewcommand{\thefigure}{F\arabic{figure}}
\setcounter{table}{0}
\setcounter{figure}{0}

\input{tables/t_tests_between}

\input{tables/t_tests_chatgpt_free_between}

%% file: tables/Sample_Description_Table.tex
\begin{table}[htb]
\centering
\begin{minipage}{0.9\textwidth}
\centering
\caption{\textbf{Demographics comparison: total sample vs. subsample (both systems users)}}
\label{tab:sample_demogra_table}
\small

\begin{tabular}{llrr}
\toprule
\multicolumn{2}{c}{} & \textbf{Total Sample (N = 526)} & \textbf{Subsample (n = 116)} \\
\multicolumn{2}{c}{} & \multicolumn{2}{r}{\footnotesize$n$ (\%)} \\
\midrule

\multicolumn{4}{l}{\textbf{User Group}} \\
 & Students              & 262 (49.9\%) & 22 (19.1\%) \\
 & Research staff        & 146 (27.8\%) & 52 (45.2\%) \\
 & Administrative staff  & 89 (17.0\%)  & 35 (30.4\%) \\
 & Professors            & 21 (4.0\%)   & 4 (3.5\%) \\
 & Other                 & 7 (1.3\%)    & 2 (1.7\%) \\

\addlinespace
\multicolumn{4}{l}{\textbf{Gender}} \\
 & Male              & 250 (47.5\%) & 63 (54.3\%) \\
 & Female            & 246 (46.8\%) & 46 (39.7\%) \\
 & Diverse           & 10 (1.9\%)   & 1 (0.9\%) \\
 & Other             & 1 (0.2\%)    & 0 (0.0\%) \\
 & Prefer not to say & 18 (3.4\%)   & 6 (5.2\%) \\

\addlinespace
\multicolumn{4}{l}{\textbf{Age}} \\
 & Under 25     & 152 (28.9\%) & 14 (12.1\%) \\
 & 25--34       & 219 (41.6\%) & 53 (45.7\%) \\
 & 35--44       & 80 (15.2\%)  & 31 (26.7\%) \\
 & 45--54       & 34 (6.5\%)   & 9 (7.8\%) \\
 & 55--64       & 38 (7.2\%)   & 9 (7.8\%) \\
 & 65 or older  & 2 (0.4\%)    & 0 (0.0\%) \\

\bottomrule
\end{tabular}
\end{minipage}
\end{table}

%% file: tables/table_chatGPT_plan.tex
\begin{table}[htb]
\caption{\textbf{ChatGPT plan usage: All ChatGPT users vs. subsample (both systems users).}}
\label{tab:chatgpt_plan}
\small

\begin{minipage}{0.85\textwidth}
\centering

\begin{tabular}{llrr}
\toprule
\multicolumn{2}{c}{} & \textbf{All ChatGPT Users (n = 333)} & \textbf{Subsample (n = 116)} \\
\multicolumn{2}{c}{} & \multicolumn{2}{r}{\footnotesize$n$ (\%)} \\
\midrule
\multicolumn{4}{l}{\textbf{ChatGPT Plan}} \\
 & Paid version     & 53 (15.9\%)  & 12 (10.3\%) \\
 & Free version     & 277 (83.2\%) & 101 (87.1\%) \\
 & Not specified    & 3 (0.9\%)    & 3 (2.6\%) \\
\bottomrule
\end{tabular}

\end{minipage}
\end{table}

%% file: tables/Correlations_CustomizedLLMaaS.tex
\begin{table}[htbp]
\caption{\textbf{Means, standard deviations, Cronbach's $\alpha$, and correlations with confidence intervals for the customized LLMaaS chatbot (subsample of both systems users, $n = 116$).}}
\small
\centering
\setlength{\tabcolsep}{2.5pt}
\begin{tabular}{lcccccccccc}
\toprule
Variable & M & SD & $\alpha$ & 1 & 2 & 3 & 4 & 5 & 6 & 7 \\
\midrule
1. Trust & 3.60 & 0.72 & 0.59 & -- & & & & & & \\
2. Cautious behavior toward hallucinations & 4.18 & 0.68 & 0.40 & \makecell{-0.06 \\ {[-0.24,0.13]}} & -- & & & & & \\
3. Perceived hallucinations & 3.23 & 1.00 & 0.82 & \makecell{-0.36** \\ {[-0.51,-0.19]}} & \makecell{0.18 \\ {[-0.01,0.35]}} & -- & & & & \\
4. Privacy concerns & 2.38 & 0.95 & 0.89 & \makecell{-0.48** \\ {[-0.61,-0.32]}} & \makecell{-0.05 \\ {[-0.23,0.13]}} & \makecell{0.31** \\ {[0.13,0.47]}} & -- & & & \\
5. Sustainability-aware AI use & 2.33 & 1.18 & 0.79 & \makecell{0.03 \\ {[-0.16,0.21]}} & \makecell{0.04 \\ {[-0.14,0.22]}} & \makecell{-0.01 \\ {[-0.19,0.17]}} & \makecell{0.21* \\ {[0.03,0.38]}} & -- & & \\
6. Perceived usefulness & 4.01 & 0.71 & 0.86 & \makecell{0.46** \\ {[0.31,0.59]}} & \makecell{-0.07 \\ {[-0.25,0.11]}} & \makecell{-0.19* \\ {[-0.36,0.00]}} & \makecell{-0.28** \\ {[-0.44,-0.10]}} & \makecell{-0.03 \\ {[-0.21,0.15]}} & -- & \\
7. Perceived ease of use & 4.59 & 0.65 & 0.91 & \makecell{0.33** \\ {[0.15,0.48]}} & \makecell{-0.00 \\ {[-0.19,0.18]}} & \makecell{-0.03 \\ {[-0.21,0.15]}} & \makecell{-0.23* \\ {[-0.40,-0.05]}} & \makecell{-0.05 \\ {[-0.23,0.14]}} & \makecell{0.28** \\ {[0.10,0.44]}} & -- \\
8. Trust in the university & 3.91 & 0.80 & 0.90 & \makecell{0.38** \\ {[0.21,0.52]}} & \makecell{-0.04 \\ {[-0.22,0.14]}} & \makecell{-0.21* \\ {[-0.37,-0.02]}} & \makecell{-0.19* \\ {[-0.36,-0.01]}} & \makecell{-0.00 \\ {[-0.19,0.18]}} & \makecell{0.33** \\ {[0.15,0.48]}} & \makecell{0.23* \\ {[0.05,0.40]}} \\
\bottomrule
\multicolumn{11}{l}{\footnotesize Notes. $\alpha$ = Cronbach’s Alpha. Values in brackets = 95\% CI. * $p < .05$, ** $p < .01$.}
\end{tabular}
\label{tab:llmaas_correlations}
\end{table}

%% file: tables/Correlations_chatGPT.tex
\begin{table}[htbp]
\caption{\textbf{Means, standard deviations, Cronbach's $\alpha$, and correlations with confidence intervals for ChatGPT (subsample of both systems users, $n = 116$).}}
\small
\centering
\setlength{\tabcolsep}{2.5pt}
\begin{tabular}{lcccccccccc}
\toprule
Variable & M & SD & $\alpha$ & 1 & 2 & 3 & 4 & 5 & 6 & 7 \\
\midrule
1. Trust & 3.08 & 0.79 & 0.71 & -- & & & & & & \\
2. Cautious behavior toward hallucinations & 4.13 & 0.67 & 0.40 & \makecell{-0.05 \\ {[-0.23,0.13]}} & -- & & & & & \\
3. Perceived hallucinations & 3.85 & 0.85 & 0.79 & \makecell{-0.23* \\ {[-0.39,-0.05]}} & \makecell{0.32** \\ {[0.15,0.48]}} & -- & & & & \\
4. Privacy concerns & 3.61 & 0.85 & 0.84 & \makecell{-0.35** \\ {[-0.50,-0.18]}} & \makecell{0.00 \\ {[-0.18,0.18]}} & \makecell{0.29** \\ {[0.12,0.45]}} & -- & & & \\
5. Sustainability-aware AI use & 2.22 & 1.17 & 0.75 & \makecell{0.10 \\ {[-0.08,0.28]}} & \makecell{-0.02 \\ {[-0.20,0.17]}} & \makecell{-0.08 \\ {[-0.26,0.10]}} & \makecell{0.02 \\ {[-0.16,0.20]}} & -- & & \\
6. Perceived usefulness & 3.98 & 0.82 & 0.87 & \makecell{0.56** \\ {[0.42,0.67]}} & \makecell{0.07 \\ {[-0.11,0.25]}} & \makecell{0.02 \\ {[-0.17,0.20]}} & \makecell{-0.19* \\ {[-0.36,-0.01]}} & \makecell{0.00 \\ {[-0.18,0.19]}} & -- & \\
7. Perceived ease of use & 4.48 & 0.64 & 0.90 & \makecell{0.18 \\ {[-0.00,0.35]}} & \makecell{0.17 \\ {[-0.01,0.34]}} & \makecell{0.26** \\ {[0.08,0.42]}} & \makecell{0.04 \\ {[-0.14,0.22]}} & \makecell{-0.04 \\ {[-0.23,0.14]}} & \makecell{0.39** \\ {[0.22,0.53]}} & -- \\
8. Trust in the university & 3.91 & 0.80 & 0.90 & \makecell{0.27** \\ {[0.10,0.44]}} & \makecell{0.03 \\ {[-0.15,0.21]}} & \makecell{-0.11 \\ {[-0.29,0.07]}} & \makecell{-0.00 \\ {[-0.18,0.18]}} & \makecell{0.03 \\ {[-0.16,0.21]}} & \makecell{0.15 \\ {[-0.04,0.32]}} & \makecell{0.15 \\ {[-0.03,0.33]}} \\
\bottomrule
\multicolumn{11}{l}{\footnotesize Notes. $\alpha$ = Cronbach’s Alpha. Values in brackets = 95\% CI. * $p < .05$, ** $p < .01$.}
\end{tabular}
\label{tab:chatgpt_correlations}
\end{table}

%% file: tables/Knowledge_Scores_Distribution.tex
\begin{table}[htb]
\centering
\begin{minipage}{0.6\textwidth} 
\centering
\caption{\textbf{Distribution of knowledge test scores for the subsample of both systems users ($n = 116$). Scores represent the number of questions answered correctly.}}
\label{tab:knowledge_score_distribution}
\small
\setlength{\tabcolsep}{12pt}

\begin{tabular}{cc}
\toprule
\textbf{Knowledge Score (0--5)} & \textbf{$n$ (\%)} \\
\midrule
0 & 17 (14.7\%) \\
1 & 11 (9.5\%) \\
2 & 22 (19.0\%) \\
3 & 13 (11.2\%) \\
4 & 28 (24.1\%) \\
5 & 25 (21.6\%) \\
\bottomrule
\end{tabular}
\end{minipage}
\end{table}

%% file: tables/t_tests_no_switchers.tex
\begin{table}[htbp]
\centering
\captionsetup{justification=raggedright,singlelinecheck=false}
\caption{\textbf{Paired t-test comparisons of the customized LLMaaS chatbot (UniGPT) vs. ChatGPT – Non-switchers ($n = 68$)}}
\label{tab:ttest_non_switchers}
\small
\setlength{\tabcolsep}{2.5pt}

\begin{minipage}{\columnwidth}
\centering

\begin{tabular*}{\textwidth}{@{\extracolsep{\fill}}lcccccccccc}
\toprule
\textbf{Dependent Variable} & \textbf{\makecell{\textit{M}\\(UniGPT)}} & \textbf{\makecell{\textit{SD}\\(UniGPT)}} & \textbf{\makecell{\textit{M}\\(ChatGPT)}} & \textbf{\makecell{\textit{SD}\\(ChatGPT)}} & \textbf{\textit{t}} & \textbf{\textit{df}} & \textbf{\textit{p}} & \textbf{Holm $\alpha$\textsuperscript{1}} & \textbf{\textit{d}\textsuperscript{2}} & \textbf{95\% \textit{CI}\textsuperscript{3}} \\
\midrule
Privacy concerns (H5) & 2.32 & 0.89 & 3.46 & 0.80 & -9.34 & 67 & < 0.001 & 0.00833 & -1.13 & [-1.44, -0.82] \\
Perceived hallucinations (H4) & 3.10 & 1.05 & 3.86 & 0.84 & -6.41 & 67 & < 0.001 & 0.01 & -0.78 & [-1.05, -0.50] \\
Trust (H1) & 3.60 & 0.71 & 3.09 & 0.81 & 5.97 & 67 & < 0.001 & 0.0125 & 0.72 & [0.45, 0.99] \\
Sustainability-aware AI use (H6) & 2.30 & 1.13 & 2.25 & 1.16 & 0.91 & 67 & 0.37 & 0.0167 & 0.11 & [-0.13, 0.35] \\
Cautious behavior toward hallucinations (H3) & 4.20 & 0.67 & 4.19 & 0.66 & 0.10 & 67 & 0.92 & n/a\textsuperscript{4} & 0.01 & [-0.23, 0.25] \\
\midrule
Usefulness\textsuperscript{5} & 4.03 & 0.74 & 3.97 & 0.80 & 0.68 & 67 & 0.50 &  & 0.08 & [-0.16, 0.32] \\
Ease of use\textsuperscript{5} & 4.52 & 0.67 & 4.46 & 0.58 & 1.02 & 67 & 0.31 &  & 0.12 & [-0.11, 0.36] \\
\bottomrule
\end{tabular*}

\vspace{2pt}
\footnotesize
\raggedright
\textit{Notes.}
\textsuperscript{1} Holm-adjusted significance threshold per hypothesis rank (sequential correction).
\textsuperscript{2} Cohen’s \textit{d} is reported as the standardized effect size for each comparison.
\textsuperscript{3} Confidence intervals (CI) for Cohen's d are not adjusted for multiple comparisons.
\textsuperscript{4} Holm-adjusted thresholds are only assigned while p-values remain below $\alpha$; correction stops after the first non-significant test.
\textsuperscript{5} Ease of Use and Usefulness were analyzed exploratively and were not part of the Holm correction.
\par
\end{minipage}
\end{table}

%% file: tables/knowledge_as_control_variable.tex
\begin{table}[htb]
\centering
\caption{\textbf{Mixed-Effects Model of system type, knowledge scores, and their interaction (number of observations: 232; $n = 116$)}}
\label{tab:knowledge_control}
\small
\centering
\begin{tabular}{l c D{.}{.}{3.3} D{.}{.}{2.2} D{.}{.}{3.3}}
\toprule
Effect & \multicolumn{1}{c}{\textit{Estimate}} & \multicolumn{1}{c}{\textit{SE}} & \multicolumn{1}{c}{\textit{t}} & \multicolumn{1}{c}{\textit{p}} \\
\midrule
Trust: System & $0.389^{**}$ & 0.130 & 3.00 & 0.003 \\
Trust: System × Knowledge & 0.049 & 0.039 & 1.25 & 0.215 \\
CatiousBehavior: System & -0.058 & 0.084 & -0.69 & 0.490 \\
CatiousBehavior: System × Knowledge & 0.041 & 0.025 & 1.61 & 0.111 \\
PerceivedHallucinations: System & $-0.537^{***}$ & 0.154 & -3.49 & < 0.001 \\
PerceivedHallucinations: System × Knowledge & -0.027 & 0.046 & -0.60 & 0.553 \\
Privacy: System & $-1.228^{***}$ & 0.193 & -6.38 & < 0.001 \\
Privacy: System × Knowledge & -0.001 & 0.058 & -0.02 & 0.987 \\
Sustainability: System & 0.066 & 0.101 & 0.65 & 0.514 \\
Sustainability: System × Knowledge & 0.016 & 0.030 & 0.53 & 0.594 \\
\bottomrule
\multicolumn{5}{l}{\footnotesize{$^{*}p < .05$; $^{**}p < .01$; $^{***}p < .001$}} \\
\multicolumn{5}{l}{\scriptsize{All models include random intercepts for participants.}} \\
\end{tabular}
\end{table}

%% file: tables/t_tests_between.tex
\begin{table*}[htb]
\centering
\begin{minipage}{\textwidth}
\centering
\caption{\textbf{Welch's independent t-test comparisons (between-subjects) of the customized LLMaaS chatbot (UniGPT) and ChatGPT, with participants using only one system (n$_{\text{UniGPT}}$ = 70, n$_{\text{ChatGPT}}$ = 217)}}
\label{tab:ttest_between_table}
\small
\setlength{\tabcolsep}{2.5pt}

\begin{tabular*}{\textwidth}{@{\extracolsep{\fill}}lcccccccccc}
\toprule
\textbf{Dependent Variable} & \textbf{\makecell{\textit{M}\\(UniGPT)}} & \textbf{\makecell{\textit{SD}\\(UniGPT)}} & \textbf{\makecell{\textit{M}\\(ChatGPT)}} & \textbf{\makecell{\textit{SD}\\(ChatGPT)}} & \textbf{\textit{t}} & \textbf{\textit{df}} & \textbf{\textit{p}} & \textbf{Holm $\alpha$\textsuperscript{1}} & \textbf{\textit{d}\textsuperscript{2}} & \textbf{95\% \textit{CI}\textsuperscript{3}} \\
\midrule
Privacy concerns & 2.47 & 0.96 & 3.17 & 0.90 & -5.32 & 110.54 & < 0.001 & 0.0083 & -0.76 & [-1.03, -0.48] \\
Sustainability-aware AI use & 2.88 & 1.21 & 2.19 & 1.03 & 4.28 & 103.37 & < 0.001 & 0.01 & 0.64 & [0.36, 0.91] \\
Perceived hallucinations & 3.40 & 1.05 & 3.75 & 0.94 & -2.48 & 106.38 & 0.01 & 0.0125 & -0.36 & [-0.63, -0.09] \\
Cautious behavior toward hallucinations & 4.29 & 0.66 & 4.10 & 0.74 & 1.95 & 129.60 & 0.05 & 0.0167 & 0.25 & [-0.02, 0.52] \\
Trust & 3.33 & 0.77 & 3.27 & 0.75 & 0.54 & 113.91 & 0.59 & n/a\textsuperscript{4} & 0.07 & [-0.19, 0.34] \\
\midrule
Usefulness\textsuperscript{5} & 3.61 & 0.88 & 3.83 & 0.82 & -1.84 & 109.78 & 0.07 &  & -0.26 & [-0.53, 0.01] \\
Ease of use\textsuperscript{5} & 4.46 & 0.66 & 4.60 & 0.59 & -1.63 & 105.83 & 0.11 &  & -0.24 & [-0.51, 0.03] \\
\bottomrule
\end{tabular*}

\vspace{2pt}
\footnotesize
\raggedright
\textit{Notes.}
\textsuperscript{1} Holm-adjusted significance threshold per hypothesis rank (sequential correction).
\textsuperscript{2} Cohen’s \textit{d} is reported as the standardized effect size for each comparison.
\textsuperscript{3} Confidence intervals (CI) for Cohen's d are not adjusted for multiple comparisons.
\textsuperscript{4} Holm-adjusted thresholds are only assigned while p-values remain below $\alpha$; correction stops after the first non-significant test.
\textsuperscript{5} Ease of Use and Usefulness were analyzed exploratively and were not part of the Holm correction.
\par
\end{minipage}
\end{table*}

%% file: tables/t_tests_chatgpt_free_between.tex
\begin{table*}[htb]
\centering
\begin{minipage}{\textwidth}
\centering
\caption{\textbf{Welch's independent t-test comparisons (between-subjects) of the customized LLMaaS chatbot (UniGPT) and ChatGPT Free, with participants using only one system ($n_{\text{UniGPT}} = 70$, $n_{\text{ChatGPT Free}} = 176$)}}
\label{tab:ttest_between_free}
\small
\setlength{\tabcolsep}{2.5pt}

\begin{tabular*}{\textwidth}{@{\extracolsep{\fill}}lcccccccccc}
\toprule
\textbf{Dependent Variable} & \textbf{\makecell{\textit{M}\\(UniGPT)}} & \textbf{\makecell{\textit{SD}\\(UniGPT)}} & \textbf{\makecell{\textit{M}\\(ChatGPT)}} & \textbf{\makecell{\textit{SD}\\(ChatGPT)}} & \textbf{\textit{t}} & \textbf{\textit{df}} & \textbf{\textit{p}} & \textbf{Holm $\alpha$\textsuperscript{1}} & \textbf{\textit{d}\textsuperscript{2}} & \textbf{95\% \textit{CI}\textsuperscript{3}} \\
\midrule
Privacy concerns (H5) & 2.47 & 0.96 & 3.17 & 0.93 & -5.14 & 122.70 & < 0.001 & 0.0083 & -0.74 & [-1.02, -0.45] \\
Sustainability-aware AI use (H6) & 2.88 & 1.21 & 2.19 & 1.05 & 4.19 & 112.56 & < 0.001 & 0.01 & 0.63 & [0.35, 0.91] \\
Perceived hallucinations (H4) & 3.40 & 1.05 & 3.77 & 0.98 & -2.56 & 118.90 & 0.011 & 0.0125 & -0.37 & [-0.65, -0.09] \\
Cautious behavior toward hallucinations (H3) & 4.29 & 0.66 & 4.15 & 0.71 & 1.39 & 136.89 & 0.17 & 0.0167 & 0.19 & [-0.09, 0.47] \\
Trust (H1) & 3.33 & 0.77 & 3.22 & 0.76 & 0.92 & 125.53 & 0.36 & n/a\textsuperscript{4} & 0.13 & [-0.15, 0.41] \\
\midrule
Usefulness\textsuperscript{5} & 3.61 & 0.88 & 3.74 & 0.84 & -1.04 & 121.41 & 0.30 &  & -0.15 & [-0.43, 0.13] \\
Ease of use\textsuperscript{5} & 4.46 & 0.66 & 4.59 & 0.59 & -1.47 & 114.81 & 0.14 &  & -0.22 & [-0.50, 0.06] \\
\bottomrule
\end{tabular*}

\vspace{2pt}
\footnotesize
\raggedright
\textit{Notes.}
\textsuperscript{1} Holm-adjusted significance threshold per hypothesis rank (sequential correction).
\textsuperscript{2} Cohen’s \textit{d} is reported as the standardized effect size for each comparison.
\textsuperscript{3} Confidence intervals (CI) for Cohen's d are not adjusted for multiple comparisons.
\textsuperscript{4} Holm-adjusted thresholds are only assigned while p-values remain below $\alpha$; correction stops after the first non-significant test.
\textsuperscript{5} Ease of Use and Usefulness were analyzed exploratively and were not part of the Holm correction.
\par
\end{minipage}
\end{table*}